\documentclass[useAMS, usenatbib, usegraphicx]{mn2e} 

\title[Galactic outflow simulated properties] 
{Galactic outflow and diffuse gas properties at $z \geq 1$ using different baryonic feedback models} 

\author[P. Barai et al.] 
{Paramita Barai$^{1}$\thanks{E-mail: pbarai@oats.inaf.it}, 
Pierluigi Monaco$^{2, 1}$, 
Giuseppe Murante$^{1}$, 
Antonio Ragagnin$^{2}$, 
\newauthor 
Matteo Viel$^{1, 3}$ 
\vspace{0.2cm} \\ 
$^{1}$ INAF - Osservatorio Astronomico di Trieste, Via G.B. Tiepolo 11, I-34143 Trieste, Italy \\ 
$^{2}$ Dipartimento di Fisica dell'Universit\`{a} di Trieste, Sezione di Astronomia, 
Via Tiepolo 11, I-34131 Trieste, Italy \\ 
$^{3}$ INFN / National Institute for Nuclear Physics, Via Valerio 2, I-34127 Trieste, Italy 
} 

\begin{document} 

\maketitle 

\label{firstpage} 

\begin{abstract} 

% which incorporates a direct distribution of thermal and kinetic energy from SN to the neighbouring gas, 
% using the free parameters of feedback efficiency fraction and a probability; 
% SN feedback in the MUPPI models is able to enrich the CGM of most galaxies to a relatively uniform value; 
% while the Effective models render some galaxies significantly less enriched. 
% Kinetic SN feedback creates more prominent hot halos 
% (consisting of outflowing and inflowing gas) in massive galaxies. 
% The simulated SFRD has a plateau maximum between $z = 2 - 4$, 
% with a steeper reduction of SFRD on either side. 
% The global star formation rate density versus redshift shows that a 
% where SN feedback is implemented using fully local gas properties. % and explore the 
% including metal cooling, chemical enrichment, % including energy-driven wind feedback. 
% of star formation (SF) 
% The number fraction of galaxies possessing outflowing gas is between $(30 - 97) \%$, 
% and increases with galaxy mass and star formation rate (SFR). 

We measure and quantify properties of galactic outflows and diffuse gas at $z \geq 1$ 
in cosmological hydrodynamical simulations 
performed using the {\sc GADGET-3} code containing novel baryonic feedback models. 
Our sub-resolution model, MUPPI, implements supernova feedback using fully local gas properties, 
where the wind velocity and mass loading are not given as input. 
We find the following trends at $z = 2$ by analysing central galaxies having a stellar mass 
higher than $10^{9} M_{\odot}$. 
The outflow velocity and mass outflow rate ($\dot{M}_{\rm out}$) 
exhibit positive correlations with galaxy mass and with the star formation rate (SFR). 
However, most of the relations present a large scatter. 
The outflow mass loading factor ($\eta$) is between $0.2 - 10$, with an average $\eta \sim 1$. 
The comparison Effective model generates a constant outflow velocity 
as expected from the input fixed wind kick speed, 
and a negative correlation of $\eta$ with halo mass as opposed to the fixed input $\eta$. 
The shape of the outflows is bi-polar in $95 \%$ of the MUPPI galaxies. 
The MUPPI model produces colder galaxy cores 
and flatter gas metallicity radial profiles than the Effective model. 
The number fraction of galaxies where outflow is detected decreases at lower redshifts, 
but remains more than $80 \%$ over $z = 1 - 5$. 
High SF activity at $z \sim 2 - 4$ drives strong outflows, 
causing the positive and steep correlations of velocity and $\dot{M}_{\rm out}$ with SFR. 
The outflow velocity correlation with SFR becomes flatter at $z = 1$, 
and $\eta$ displays a negative correlation with halo mass in massive galaxies. 
Our study demonstrates that both the MUPPI and Effective models 
produce significant outflows at $\sim 1 / 10$ of the virial radius; 
at the same time shows that the properties of outflows generated can be different from 
the input speed and mass loading in the Effective model. 
Our MUPPI model, {\it using local properties} of gas in the sub-resolution recipe, 
is able to develop galactic outflows whose properties {\it correlate with global galaxy properties}, 
and consistent with observations.

\end{abstract} 


\section{Introduction} 
\label{sec-intro} 

%\footnote{Here, we use the terms {\it wind} and {\it outflow}
%synonymously, meaning continuous outward flow of gas from a galaxy,
%which might or might not escape depending on its velocity and galactic potential.} 
% The feedback processes cause the baryons to cycle between galaxies and IGM, plays an imperative role. 
% These outflows are also argued to affect the large column density parts of 
% the Lyman-$\alpha$ absorption line forest (LLSs, subDLAs and DLAs) 
% seen in the spectra of distant quasars, which traces the IGM matter distribution 
% \citep[e.g.,][]{McDonald05, Kollmeier06, Tescari11, Viel12}. 
% and their interplay with the larger-scale environments is investigated. 
% If we consider the underlying physics, 
% which occurs on scales orders of magnitude below the scales resolved in cosmological simulations, 
% such an outflow is likely to be driven by the thermal pressure of SN, 
% and might be (more physically) called a {\it thermally-driven wind}. 

% using analytical prescriptions from \citet{Murray05} for
% momentum injection provided by radiation pressure of photons and SN.
% few constant-mass loading and constant-velocity cases; mass loading and velocity dependent on gas density,
% gravitational potential, and halo circular velocity; 
% wind particles temporarily decoupled hydrodynamically versus not decoupled; 

% Sub-pc scale simulations \citep[e.g.,][]{Federrath14} are needed to study 
% the actual jet and outflow feedback associated with SF. 

Baryons existing in cosmic structures 
regularly undergo the processes of star formation (SF) and supernovae (SN) explosions, 
which subsequently deposit a fraction of mass/energy to the surroundings. 
This energy feedback heats up, ionizes and drives gas outward, 
often generating large-scale {\it outflows/winds} 
\citep[e.g.,][]{Burke68, Mathews71, Vader86, Veilleux05, Rubin10}. 
Galactic outflows are observed at low redshifts 
\citep[e.g.,][]{Burbidge64, Fabbiano84, Ohyama97, Smith05, Arribas14}, 
reaching velocity as large as $1000$ km/s \citep{Diamond-Stanic12, Bradshaw13}, 
and at high-$z$ \citep[e.g.,][]{Frye02, Wilman05, Weiner09, Kornei12, Tang14}, 
up to $z \sim 5$ \citep{Dawson02}, 
sometimes extending over distances of $60 - 130$ physical kpc \citep[e.g.,][]{Steidel11, Lundgren12}. 

Winds driven by starbursts and SN are an important source of feedback in galaxy evolution. 
They are considered to be the primary mechanism by which 
metals are ejected out of star-forming regions in galaxies and 
deposited into the circumgalactic medium (CGM) and the intergalactic medium (IGM) 
\citep[e.g.,][]{Larson75, Aguirre01, Aracil04, Fox07, Pinsonneault10, Gauthier12}. 
They constitute a key ingredient of galaxy formation models, 
both in hydrodynamic simulations \citep[e.g.,][]{Oppenheimer08, Haas13, Hirschmann13, Tescari14, Bird14}, 
and in semi-analytical models \citep[e.g.,][]{Baugh06, Benson12}. 
The outflows are argued to quench SF and suppress the formation of low-mass galaxies, 
by expelling out gas available to make stars. 
This in turn flattens the low-mass end of the simulated galaxy mass function 
\citep[e.g.,][]{Theuns02, Rasera06, Stinson07}, bringing it closer to observations. 
Feedback is also invoked to reproduce realistic disk galaxies in cosmological hydrodynamic simulations 
\citep[e.g.,][]{Weil98, Sommer-Larsen03, Okamoto05, Scannapieco12, Angles-Alcazar14}. 

The physical mechanisms for the origin and driving of galactic winds is complex, 
occurring on scales of the multiphase structure of the interstellar medium (ISM) 
and molecular clouds \citep[e.g.,][]{Heckman03}. 
The gas is likely accelerated either by thermal pressure \citep[e.g.,][]{Chevalier85}, 
radiation pressure \citep[e.g.,][]{Murray05} 
with an impact of dust \citep{Thompson14}, 
cosmic rays \citep[e.g.,][]{Vazza14}; 
or a combination of them \citep[e.g.,][]{Sharma12}. 
The relevant physical scales are orders of magnitude below 
the scales resolved in cosmological simulations. 
Hence energy ejection by starburst and SN are incorporated 
in the simulations using {\it sub-resolution} numerical prescriptions. 
The physics of the multiphase ISM, on scales unresolved in cosmological simulations, 
is modelled using spatially averaged properties describing the medium on scales that are resolved. 

Thermal feedback, where SN energy is distributed as heating energy of the neighbouring gas, 
is historically known to be ineffective \citep[e.g.,][]{Friedli95, Katz96}, 
because fast cooling of the dense SF gas radiates away the thermal energy quickly. 
A few studies have proposed numerical schemes for efficient thermal feedback in 
smoothed particle hydrodynamics (SPH) simulations \citep[e.g.,][]{Kay03, DallaVecchia12}. 
Depositing the SN energy in the kinetic form is a more popular implementation in the literature, 
which has been shown to have significant feedback effects 
\citep[e.g.,][]{Navarro93, Cen00, Kawata01, Dubois08, Oppenheimer12}. 
Some other approaches of numerical SN feedback are: 
consider that a part of the neighbouring gas undergoes adiabatic evolution 
by turning off radiative cooling temporarily 
\citep[e.g.,][]{Mori97, Thacker00, Brook05, Stinson06, Piontek11}; 
distribute SN energy to hot and cold gas phases separately \citep[e.g.,][]{Marri03, Scannapieco06}. 

Kinetic SN feedback models for cosmological simulations 
generally imparts a velocity kick to the affected gas. 
An {\it energy-driven wind} scheme was proposed by \citet{SH03}, and 
subsequently used by others \citep[e.g.,][]{Tornatore04, DallaVecchia08, Tescari09, Fabjan10, Barai13}. 
In this framework, a fraction of SN energy provides the outflow kinetic energy, 
and the wind speed is constant. 
Other models formulate the outflow velocity and mass loading factor 
in terms of galaxy global properties (mass, velocity dispersion, SFR): 
{\it momentum-driven wind} \citep[e.g.,][]{Oppenheimer08, Tescari09}, 
dependent on the local velocity dispersion of the dark matter \citep{Okamoto10}, 
multicomponent and variable velocity galactic outflow \citep{Choi11}, 
halo mass dependent energy-driven outflow \citep{Puchwein12}, 
combination of both energy-driven and momentum-driven cases \citep{Schaye10}. 
Very recently \citet{Schaye14} presented results from a stochastic implementation of thermal SN feedback, 
where galactic winds develop without imposing any input outflow velocity and mass loading factor. 

In our current work we explore the novel sub-resolution model {\bf MUPPI} 
[{\bf MU}lti-{\bf P}hase {\bf P}article {\bf I}ntegrator, \S\ref{sec-num-Muppi}] 
\citep{Murante10, Murante14}. 
It consists of a scheme of SF in multiphase ISM, where the system of ordinary differential equations 
are numerically integrated within the hydrodynamical time-step. 
MUPPI incorporates a direct distribution of thermal and kinetic energy 
from SN to the neighbouring gas, using local properties of the gas, 
and the free parameters of feedback energy efficiency fraction and a probability. 
Thus our model has no input expression of wind velocity and mass loading for SN feedback. 
Additionally our formulation, unlike widely done in the literature, 
has no dependence on global properties like galaxy mass or velocity dispersion. 
The MUPPI model has been shown to reproduce the Schmidt-Kennicutt relation \citep{Monaco12}, 
predict a warm mode of gas accretion on forming galaxies \citep{Murante12}, 
generate realistic disk galaxies at a moderate resolution \citep{Murante14}, 
and the properties of bars in spiral galaxies were studied \citep{Goz14}. 
For comparison with MUPPI, we adopt the effective SF model \citep{SH03} with two 
variations of energy-driven wind: a constant-velocity, 
and where the wind velocity depends on distance from galaxy center \citep{Barai13} 
following a correlation motivated by the observational studies of \citet{Steidel10}. 

% We argue that such models are ad-hoc, and requires the local small-scale event of SN explosion 
% to know some of the large-scale galaxy properties, which is physically impossible. 
% built on the effective SF model \citep{SH03} which used equilibrium solutions 
% for the hot and cold gas phases. Rather 
% roughly follow the input momentum-driven wind trends. 
% eventual reaccretion into SF gas is tracked. wind recycling quantified in galaxy mass & environment. 
% wind material re-accretes on to galaxy on recycling time-scale that varies inversely with galaxy mass. 
% or gas particles receiving a velocity kick 
% and detected that the recycling time varies with galaxy mass. 
% since the recycling time of winds is shorter in higher-mass galaxies 
% because of the interaction between outflows 
% and the increasingly higher gas densities in and around massive haloes; 
% this leads to increased accretion in massive galaxies, modifying the current stellar mass function. 
% and found that low-ionisation metal absorbers (e.g. MgII) are from gas 
% which will fall into galaxies within several Gyr, 
% while high-ionisation metal absorbers (e.g. OVI) trace materials that was deposited by outflows Gyr's ago. 

Studies have analysed CGM gas inflows and outflows in cosmological hydrodynamic simulations. 
\citet{Oppenheimer08} tracked wind particles (\S\ref{sec-num-EffMod}) 
and computed the feedback mass and energy outflow rate as a function of galaxy baryonic mass. 
\citet{Oppenheimer10} followed the gas previously ejected in winds 
and accreting for new SF at $z < 1$, measuring the recycling time. 
\citet{Ford13} investigated the CGM dynamical state by tracking 
inflowing, outflowing, and ambient gas based on cross-correlation of the gas particle locations 
between $z = 0.25$ and $z = 0$ with respect to galaxy positions. 
\citet{vanDeVoort12} presented gas physical property (density, temperature) radial profiles, 
for all, inflowing and outflowing gas, inside and around galaxy haloes at $z = 2$. 
Using a different set of numerical schemes in the zoom-in simulation of a massive galaxy at $z = 3$, 
\citet{Shen12} presented the time evolution of outflow velocity, metallicity, 
and the mass-loading factor for the main host and the massive dwarf galaxies. 
In similar zoom simulations, \citet{Brook13} examined 
the time evolution of gas outflow rates through $R_{\rm vir} / 8$, and their metallicities. 
Performing 3D hydrodynamic simulations of a disc-halo system on 10's-pc scale, 
\citet{vonGlasow13} injected SN energy as superbubbles, and studied the development of bipolar outflows, 
together with their mass and energy loss rates. 

In this paper we perform a quantification of SN-driven galactic outflow properties 
in cosmological hydrodynamical simulations. 
As described earlier, outflows as widely observed in galaxies, 
are generated in simulations using thermal or (more popularly) kinetic SN feedback. 
In several cases a pre-defined wind velocity and mass loading factor are input to the sub-resolution model. 
However the velocity and mass loading which are imparted to the gas in a numerical scheme 
might not be the speed and rate with which an outflow actually develops. 
This is because the outflow propagates through the atmosphere of a galaxy and 
interacts with surrounding gas. 
Therefore determining these properties explicitly is necessary 
for a complete study of simulated galactic outflows, which we do in this paper. 
Furthermore the MUPPI model has no input expression of wind velocity and mass loading for SN feedback. 
We compute the velocity, mass outflow rate and mass loading of our simulated galaxies, 
and investigate how differently MUPPI drives outflows than other models. 
We measure the characteristics of outflows of a statistical sample of galaxies 
extracted from cosmological simulations, over redshifts $z = 1 - 5$, 
aiming to infer possible correlations with galaxy mass and SFR. 
We also study the properties of diffuse gas in the CGM of galaxies. 

This paper is organised as follows: 
we describe our numerical code in \S\ref{sec-numerical}, 
the simulations in \S\ref{sec-num-sim}, 
present and discuss our results in \S\ref{sec-results}, and 
finally give a summary of the main findings in \S\ref{sec-conclusion}.

\section{Numerical Method} 
\label{sec-numerical} 

We use a modified version of the TreePM (particle mesh) - 
SPH (smoothed particle hydrodynamics) code {\sc GADGET-3}, 
which includes a more efficient domain decomposition 
to improve the work-load balance over {\sc GADGET-2} \citep{Springel05}. 
The additional sub-resolution\footnote{{\it sub-resolution} refers to 
processes occurring at physical scales smaller than the resolved scales in our simulations.} 
physics included in our code are outlined below. 
There is no active galactic nuclei (AGN) feedback in our models.

\subsection{Radiative Cooling} 
\label{sec-num-cool} 

Radiative cooling is implemented by adopting the cooling rates 
from the tables of \citet{Wiersma09a}, which includes metal-line cooling. 
Eleven element species (H, He, C, Ca, O, N, Ne, Mg, S, Si, Fe) 
are tracked for constructing the cooling tables. 
Heating from a spatially-uniform time-dependent photoionizing radiation is considered 
from the cosmic microwave background and 
the \citet{Haardt01} model for the ultraviolet/X-ray background produced by quasars and galaxies. 
The gas is assumed to be dust-free, optically thin and in (photo-)ionization equilibrium. 
Contributions from the 11 elements are interpolated as a function of density, temperature and redshift 
from tables that have been pre-computed using the 
public photoionization code CLOUDY \citep[last described by][]{Ferland98}. 
Note that one simulation ({\it M50std}, defined in Table~\ref{Table-Sims}) 
is done using cooling rates from the tables of \citet{Sutherland93}, 
which tracks eight elements (H, He, C, O, Mg, S, Si, Fe). 
In this case, the relative abundance of various metals in gas is fixed to Solar, 
and only the overall metallicity counts to computing the cooling rates.

\subsection{Chemical Evolution} 
\label{sec-num-Chemistry} 

Stellar evolution and chemical enrichment are followed according to the model 
introduced by \citet{Tornatore07}. 
Each star particle is treated as a simple stellar population (SSP). 
Given a stellar initial mass function (IMF), the mass of the SSP is varied in time 
following the death of stars, and accounting for stellar mass losses. 
Production of 9 metal species (C, Ca, O, N, Ne, Mg, S, Si, Fe) is accounted for using yields from 
Type Ia SN \citep{Thielemann03}, Type II SN \citep{Woosley95}, 
as well as asymptotic giant branch stars \citep{vandenHoek97}. 
Different stellar populations release metals with mass-dependent time delays, 
where the lifetime function by \citet{Padovani93} is adopted. 
The mass range for SN-II is $M / M_{\odot} > 8$, 
while that for SN-Ia originating from binary systems is $0.8 < M / M_{\odot} < 8$ 
with a binary fraction of $10\%$. 
SN-Ia (present only in the effective model - \S\ref{sec-num-EffMod}, not in MUPPI - \S\ref{sec-num-Muppi}) 
and SN-II contribute to thermal energy feedback. 

The seven new simulations presented in this paper (\S\ref{sec-num-sim}) assume a fixed stellar IMF 
following \cite{Kroupa93}, in the mass range $(0.1 - 100) M_{\odot}$. 
Two older simulations (used for comparison) include an IMF by \citet{Chabrier03}. 
The two IMFs are similar ​in the mass range considered, 
so we assume they do not cause any difference in our outflow analysis results. 
Stars within a mass interval $[8 - 40] M_{\odot}$ become SN first 
before turning into black holes (BHs) at the end of their lives, while 
stars of mass $> 40 M_{\odot}$ are allowed to directly end in BHs without contributing to enrichment. 
Here by BH we mean stellar-mass BHs, whose evolution is not followed; 
they enter only in the stellar evolution prescriptions. 

The chemical evolution model also incorporates mass loss through stellar winds and SN explosions, 
which are self-consistently computed for a given IMF and lifetime function. 
A fraction of a star particle's mass is restored as diffuse gas during its evolution, 
and distributed to the surrounding gas. 
The enriched material is spread among the neighbouring gas particles with weights given by the SPH kernel. 

% in the thermally pulsating asymptotic giant branch (TP-AGB) phase. 
% which is a power-law at $M / M_{\odot} > 1$ and has a log-normal form at masses below. 
% However, we use power-law IMFs with different slopes 
% over the whole mass range of $0.1$ to $100 M_{\odot}$, 
% which has been tested to mimic the log-normal form of \citet{Chabrier03} at lower masses. 
% The functional form: $\phi \left( M \right) = K M^{-y}$, 
% is composed of 3 slopes and normalizations in our model: 
% $y = 0.2$ and $K = 0.497$ for stellar masses $0.1 \leq M / M_{\odot} < 0.3$, 
% $y = 0.8$ and $K = 0.241$ for $0.3 \leq M / M_{\odot} < 1$, and 
% $y = 1.3$ and $K = 0.241$ for $1 \leq M / M_{\odot} < 100$. 

\subsection{Effective Model of SF and SN Feedback} 
\label{sec-num-EffMod} 

In this model, star formation follows the effective sub-resolution scheme by \citet{SH03}. 
Gas particles with density above a limiting threshold, 
$n_{\rm SF} = 0.13$ cm$^{-3}$ (in units of number density of hydrogen atoms), are considered 
to contain two phases: cold condensed clouds, and ambient hot gas, in pressure equilibrium. 
Each gas particle represents a region of the ISM, 
where the cold clouds supply the material available for SF. 
The star forming gas remains in self-regulated equilibrium. 
Star particles are collisionless, and are spawned from gas particles undergoing SF,
according to the stochastic scheme introduced by \citet{Katz96}. 
We allow a gas particle to spawn up to four generations of star particles. 

Kinetic feedback from SN is included following the {\it energy-driven} outflow 
prescription (originally from SH03). 
The wind mass-loss rate ($\dot{M}_w$) relates to the SF rate ($\dot{M}_{\star}$) 
via the mass loading factor ($\eta$), 
\begin{equation} 
\label{eq-MdotWind} 
\dot{M}_w = \eta \dot{M}_{\star}. 
\end{equation} 
Observations reveal that mass outflow rates in galaxies are comparable to 
or a few times larger than their SF rates \citep[e.g.,][]{Martin99, Pettini02, Bouche12, Newman12a}. 
Thus, following SH03, we adopt a constant $\eta = 2$. 

The wind kinetic energy is a fixed fraction $\chi$ of SN energy: 
\begin{equation} 
\label{eq-EnrgEquate} 
\frac{1}{2} \dot{M}_w v_w^2 = \chi \epsilon_{SN} \dot{M}_{\star}. 
\end{equation} 
Here $v_w$ is the wind velocity, 
$\epsilon_{SN}$ is the average energy released by SN for each $M_{\odot}$ 
of stars formed under the instantaneous recycling approximation. 
Combining Eqs. (\ref{eq-MdotWind}) and (\ref{eq-EnrgEquate}), 
$v_w$ can be re-written as: 
\begin{equation} 
\label{eq-vW-EnrgDr} 
v_w = \left( \frac{2 \chi \epsilon_{SN}}{\eta} \right)^{1/2}. 
\end{equation} 
Following a series of studies \citep[e.g.,][]{Tornatore07, Tescari11, Barai13}, and unlike SH03, 
we choose $v_w$ as a free parameter. 
For our adopted Chabrier power-law IMF (\S\ref{sec-num-Chemistry}), 
$\epsilon_{SN} = 1.1 \times 10^{49}$ erg $M_{\odot}^{-1}$. 
% We initialize the wind energy fraction using $v_w = 400$ km/s, 
% which is the constant velocity of our simulation run CW (\S\ref{sec-num-sim}). 
% It corresponds to $\chi = 0.29$ of SN energy being carried away by the wind. 

We explore two different outflow models: constant-$v_w$ and radially-varying $v_w$. 
In the latter, developed in \citet{Barai13}, 
the wind velocity depends on galactocentric radius ($r$, or distance from galaxy center), 
following a correlation motivated by the observational studies of \citet{Steidel10}: 
\begin{equation} 
\label{eq-vSteidel} 
v_w(r) = v_{\rm max} \left( \frac{r_{\rm min}^{1-\alpha} - r^{1-\alpha}} 
			     {r_{\rm min}^{1-\alpha} - R_{\rm eff}^{1-\alpha}} \right)^{0.5}. 
\end{equation} 
Here $r_{\rm min}$ is the distance from which the wind is launched and where the velocity is zero, 
$R_{\rm eff}$ represents the outer edge of gas distribution,  
$v_{\rm max}$ is the velocity at $R_{\rm eff}$ and $\alpha$ is a power-law index. 
We choose the following parameters for our simulation run {\it E35rvw} (\S\ref{sec-num-sim}): 
$r_{\rm min} = 1 h^{-1}$ kpc, $R_{\rm eff} = 100 h^{-1}$ kpc, $v_{\rm max} = 800$ km/s, and $\alpha = 1.15$. 

% therefore a typical star particle mass is about one-fourth of the initial mass of gas particles. 
% where the mass and energy carried away by outflows are regulated by two equations. 
% from which the effective $\chi$ can be computed using Eq. (\ref{eq-vW-EnrgDr}). 

% Outflow models implemented in the {\sc GADGET} code generally involve different 
% scaling relations of $v_w$ and $\eta$ in terms of galaxy velocity dispersion, mass, and/or SFR. 
% For the original energy-driven outflows, $v_w$ and $\eta$ are constant. 
% The momentum-driven wind prescription by \citet{Oppenheimer06,Oppe08} has 
% $v_w \propto \sigma \sqrt{(L/L_{\rm crit}) - 1}$ and $\eta \propto 1 / \sigma$, 
% where $\sigma$ is the galaxy velocity dispersion, and 
% $L / L_{\rm crit}$ is its luminosity in units of a critical value. 
% In \citet{Choi11}'s model, 
% $v_w \propto {\rm SFR}^{1/3}$ and $\eta$ is a function of galaxy stellar mass. 
% \citet{Puchwein12b} assumed $v_w$ and $\eta$ to be proportional to halo mass. 

% Studying the metal-enriched gas kinematics in a region of $\sim 125$ kpc around 
% star-forming (Lyman-break) galaxies at redshifts $z = 2 - 3$, they are able to 
% reproduce their spectroscopic data using a simple model for outflows and circumgalactic gas. 

\subsection{MUPPI Model of SF and SN Feedback} 
\label{sec-num-Muppi} 

The original MUPPI sub-resolution algorithm is described in \citet{Murante10}, 
and its latest features (chemical evolution, metal cooling, SN kinetic feedback) in \citet{Murante14}. 
Gas particles undergoing SF are assumed 
(following \citealt{Monaco04}, originally from \citealt{SH03}) 
to represent a multiphase ISM composed of hot and cold phases, in pressure equilibrium. 
A fraction of the cold phase (the molecular gas fraction) provides the reservoir for SF. 
Star particles are created by a stochastic algorithm (as in \S\ref{sec-num-EffMod}). 
The hot phase is heated by the energy from massive and dying stars and radiatively cools. 

A gas particle enters the multiphase regime whenever its temperature drops below $10^5$ K, 
and its density is higher than a threshold $\rho_{\rm thr}$ 
(not to be confused with the SF density threshold $n_{\rm SF}$ in \S\ref{sec-num-EffMod}). 
A multiphase particle's evolution is governed by four ordinary differential equations (ODE) in terms 
of masses of three (hot, cold, stellar) components and thermal energy of the hot phase. 
Matter flows among the three components: 
cooling deposits hot gas into the cold phase; evaporation brings cold gas back to the hot phase; 
SF moves mass from the cold gas to stars; 
restoration moves mass from stars back to the hot phase. 
Within each SPH timestep, 
the ODEs are integrated using a Runge-Kutta method with adaptive timesteps, 
making the integration timestep much shorter than the SPH one. 
A multiphase cycle can last upto time $t_{\rm clock}$, 
set proportional to the dynamical time of the cold phase. 
A gas particle is no longer multiphase when its density reaches below $\rho_{\rm th} / 5$. 
Moreover, at low densities if SN energy is not sufficient to sustain a hot phase 
(rendering a hot temperature below $10^5$ K), the particle is forced to exit the multiphase regime. 
Opposed to the Effective Model, no equilibrium hypothesis is imposed in MUPPI. 
Here propoerties of the ISM interacts with the 
varying hydrodynamical propoerties obtained from the gas evolution. 

Energy from SN is distributed in both thermal and kinetic forms. 
The total thermal energy ejected by each star-forming multiphase particle is 
\begin{equation} 
\Delta E_{\rm heat,o} = E_{\rm SN} \cdot f_{\rm fb,out} \cdot \frac{\Delta M_\star}{M_{\rm\star, SN}} . 
\end{equation} 
Here $E_{\rm SN}$ is the energy of a single supernova, 
and $M_{\rm\star,SN}$ the stellar mass associated to each SN event. 
The thermal energy is deposited to gas neighbours within the particle's SPH smoothing length, 
and lying inside a cone of axis along the direction of minus the local density gradient 
and having a semi-aperture angle $\theta = 60^{\circ}$. 
This mimics the blowout of superbubbles along the path of least resistance \citep[see][]{Monaco04}. 
Energy contributions are weighted by the SPH kernel, using the distance from the cone axis. 
This thermal feedback scheme is relatively effective even at high densities. 

For kinetic SN feedback, when a gas particle exits a multiphase cycle, 
a probability $P_{\rm kin}$ to become a ``wind particle'' is assigned to it. 
The wind particles can receive kinetic energy from neighbouring multiphase particles 
for a time $t_{\rm wind}$. 
Assuming that outflows are driven by SNII exploding from molecular cloud destruction, 
this time is set equal to the lifetime of an $8 M_{\odot}$ star, $t_8$, 
minus the duration $t_{\rm clock}$ of the past multiphase cycle: 
\begin{equation} 
\label{eq-Muppi-twind} 
t_{\rm wind} = t_8 - t_{\rm clock} \, . 
\end{equation} 
The wind phase quits earlier than $t_{\rm wind}$ whenever the wind (gas) particle 
falls to a density below $0.3 \rho_{\rm thr}$. 
For each star-forming particle, the available kinetic energy is 
\begin{equation} 
\label{eq-Muppi-Ekin} 
E_{\rm kin}=f_{\rm fb,kin} E_{\rm SN} \, . 
\end{equation} 
The kinetic energy is distributed (using the same scheme as thermal energy) 
to wind particles inside the SPH kernel of the multiphase particle 
and within a cone of semi-aperture $\theta$ anti-aligned with the density gradient, 
and the energy contribution is weighted by the distance from the cone axis. 
The eligible wind particles receive ``velocity kicks'' as follows. 
For each wind particle, the energy contribution from all kicking particles is computed, 
and the energy-weighted average vector from kicking particles to the wind one. 
Then the kinetic energy of the wind particle is increased\footnote{In 
the reference frame of the particle itself.}, 
with the velocity increase in the direction defined above. 
In order to avoid hydrodynamical coupling at $\sim$kpc scale, 
a wind particle is decoupled from its surrounding gas as long as it remains in the wind phase. 

The free parameters of the SN feedback prescription are 
$f_{\rm fb,out}$, $f_{\rm fb,kin}$ and $P_{\rm kin}$; 
the values we explore are given in Table~\ref{Table-Sims} and \S\ref{sec-num-sim}. 
At variance with other kinetic wind models in the literature, 
neither the outflow velocity, nor the mass-loading are given as input quantities. 
Nevertheless, typical values of these can be estimated theoretically \citep{Murante14}. 
With the default parameters, the mass load factor estimate is 
\begin{equation} 
\label{eq-Muppi-eta} 
\langle \eta \rangle = \frac{ \dot{M_{\rm wind}} } { \dot{M_{\rm sfr}} } 
     = {P_{\rm kin} \over \langle f_{\rm cold} \rangle \langle f_{\rm mol} \rangle f_*} \simeq 1.5, 
\end{equation} 
and the mass-weighted average wind velocity 
\begin{equation} 
\label{eq-Muppi-Vwind2} 
\langle v_{\rm wind} \rangle = 
\sqrt{ \frac{f_{\rm fb,kin}}{\langle \eta \rangle} \frac{E_{\rm SN}}{\langle M_{*,SN}\rangle} } 
\simeq 600 ~ {\rm km/s}, 
\end{equation} 
where $\langle \rangle$ indicates that average values for star-forming ISM are used. 
Furthermore, the MUPPI feedback implementation uses only local properties of the gas.

\section{Simulations} 
\label{sec-num-sim} 

Our series of simulations is listed in Table~\ref{Table-Sims}. 
Cosmological volumes are evolved, with periodic boundary conditions, 
starting from an equal number of dark matter (DM) and gas particles at $z = 99$, up to $z = 0$. 
The initial conditions have been generated using the 
{\sc CAMB}\footnote{http://camb.info/} software \citep{lewisetal}. 
A concordance flat $\Lambda$CDM model is used, 
in agreement with recent observations of the cosmic microwave background radiation, 
weak gravitational lensing, Lyman-$\alpha$ forest and galaxy cluster mass function evolution 
\citep[e.g.,][]{Lesgourgues07, Vikhlinin09, Komatsu11}. 
The seven new runs use the following parameters: 
$\Omega_{M, 0} = 0.24, \Omega_{\Lambda, 0} = 0.76, \Omega_{B, 0} = 0.04, 
H_{0} = 72$ km s$^{-1}$ Mpc$^{-1}$. 
The two old runs \citep{Barai13} use: 
$\Omega_{M, 0} = 0.2711, \Omega_{\Lambda, 0} = 0.7289, \Omega_{B, 0} = 0.0463, 
H_{0} = 70.3$ km s$^{-1}$ Mpc$^{-1}$. 
We assume that this slight change in cosmological parameters do not cause any difference in the results. 

% $n_{\rm S} = 0.96, \sigma_8 = 0.809$
% This work describes simulation results at $z \geq 2$, 
% the analysis at $z = 0$ will be presented in a forthcoming paper. 

%%%%%%%%%%%%%%%%%%%%%%%%%%%%%%%%%%%%%%%%%%%%%%%%%%%%%%%%%%%%%%%%%%%%%%%%%%%%%%%%%%%%%%%%%% 
%%%%%%%%%%%%%%%%%%%%%%%%%%%%%%%%%%%%%%%%%%%%%%%%%%%%%%%%%%%%%%%%%%%%%%%%%%%%%%%%%%%%%%%%%% 
%
% TABLE 1 

\begin{table*} 
\begin{minipage}{16cm} 
\caption{ 
Simulation parameters. 
Column 1: Name of simulation run. 
Column 2: $L_{\rm box}$ = Comoving side of cubic simulation volume. 
Column 3: Total number of gas and DM particles in the initial condition. 
Column 4: Mass of gas particle (which has not undergone any star-formation). 
Column 5: Mass of star particle. 
Column 6: Gravitational softening length (of all particle types). 
Column 7: Model of SF and SN feedback. 
Column 8: $v_w$ = Wind velocity. 
Column 9: $f_{\rm fb,out}$ = Thermal feedback energy fraction. 
Column 10: $f_{\rm fb,kin}$ = Kinetic feedback energy fraction. 
Column 11: $P_{\rm kin}$ = Probability of kicking gas particles into wind. 
} 

\label{Table-Sims} 
\begin{tabular}{@{}ccccccccccc} 

\hline 

Run  & $L_{\rm box}$ & $N_{\rm part}$ & $m_{\rm gas}$ & $m_{\star}$ & $L_{\rm soft}$ & \multicolumn{5}{c}{SF \& SN feedback sub-resolution physics} \\ 
Name & [Mpc] & & [$M_{\odot}$] & [$M_{\odot}$] & [kpc] & Model & $v_w$ & $f_{\rm fb,out}$ & $f_{\rm fb,kin}$ & $P_{\rm kin}$ \\ 

\hline 

{\it E35nw}  & $35.56$ & $2 \times 320^3$ & $8.72 \times 10^{6}$ & $2.18 \times 10^{6}$ & $2.77$ (comoving) & Effective & $0$ \\   % EffMod25-NW 

{\it E35rvw} & $35.56$ & $2 \times 320^3$ & $8.72 \times 10^{6}$ & $2.18 \times 10^{6}$ & $2.77$ (comoving) & Effective & $v_w(r)$ \\   % EffMod25-RVW 

{\it E25cw}  & $25$ & $2 \times 256^3$ & $5.36 \times 10^{6}$ & $1.34 \times 10^{6}$ & $0.69$ (physical) & Effective & $350$ \\   % EffMod18 

{\it M25std} & $25$ & $2 \times 256^3$ & $5.36 \times 10^{6}$ & $1.34 \times 10^{6}$ & $0.69$ (physical) & MUPPI & & $0.2$ & $0.6$ & $0.03$ \\   % Muppi18 

{\it M25a} & $25$ & $2 \times 256^3$ & $5.36 \times 10^{6}$ & $1.34 \times 10^{6}$ & $0.69$ (physical) & MUPPI & & $0.4$ & $0.4$ & $0.03$ \\   % Muppi18-Fth04fk04 

{\it M25b} & $25$ & $2 \times 256^3$ & $5.36 \times 10^{6}$ & $1.34 \times 10^{6}$ & $0.69$ (physical) & MUPPI & & $0.2$ & $0.8$ & $0.03$ \\   % Muppi18-Fth02fk08 

{\it M25c} & $25$ & $2 \times 256^3$ & $5.36 \times 10^{6}$ & $1.34 \times 10^{6}$ & $0.69$ (physical) & MUPPI & & $0.2$ & $0.6$ & $0.01$ \\   % Muppi18-onepercent 

{\it M25d} & $25$ & $2 \times 256^3$ & $5.36 \times 10^{6}$ & $1.34 \times 10^{6}$ & $0.69$ (physical) & MUPPI & & $0.2$ & $0.6$ & $0.06$ \\   % Muppi18-sixpercent 

{\it M50std} & $50$ & $2 \times 512^3$ & $5.36 \times 10^{6}$ & $1.34 \times 10^{6}$ & $0.69$ (physical) & MUPPI & & $0.2$ & $0.5$ & $0.03$ \\   % Muppi36 

\hline 
\end{tabular} 

\end{minipage} 
\end{table*} 

%%%%%%%%%%%%%%%%%%%%%%%%%%%%%%%%%%%%%%%%%%%%%%%%%%%%%%%%%%%%%%%%%%%%%%%%%%%%%%%%%%%%%%%%%% 
%%%%%%%%%%%%%%%%%%%%%%%%%%%%%%%%%%%%%%%%%%%%%%%%%%%%%%%%%%%%%%%%%%%%%%%%%%%%%%%%%%%%%%%%%% 

Two simulations: {\it E35nw} and {\it E35rvw}, are taken from \citet{Barai13}. 
The relevant numbers are: 
$L_{\rm box} = 25 h^{-1} = 35.56$ Mpc comoving, $N_{\rm part} = 2 \times 320^3, 
m_{\rm gas} = 8.72 \times 10^6 M_{\odot}, L_{\rm soft} = 2.77$ kpc. 
Here the minimum gas smoothing length is set to a fraction $0.001$ of $L_{\rm soft}$. 

We perform six new runs: five varying the SN feedback parameters of Muppi, and one with the effective model. 
These are with a box-size of $L_{\rm box} = 25$ Mpc comoving, 
using $N_{\rm part} = 2 \times 256^3$ DM and gas particles in the initial condition, 
and gas particle mass $m_{\rm gas} = 5.36 \times 10^6 M_{\odot}$. 
The Plummer-equivalent softening length for gravitational forces is set to 
$L_{\rm soft} = 2.08$ kpc comoving for the evolution up to $z = 2$. 
The softening is then held fixed at $L_{\rm soft} = 0.69$ kpc in physical units from $z = 2$ to $z = 0$. 
We perform an additional Muppi run with a larger box-size of $50$ Mpc comoving, 
with the same resolution, hence using $N_{\rm part} = 2 \times 512^3$ DM and gas particles. 
The minimum gas smoothing length attainable is set to $0$ in all our seven runs. 
In all cases, the minimum smoothing 
which is actually achieved in the simulations depend on the resolution, 
and in our runs the gas smoothing lengths went down to $\sim 0.2 L_{\rm soft}$. 

All the nine simulations incorporate the metal-cooling and chemical enrichment 
sub-resolution physics described in \S\ref{sec-num-cool} and \S\ref{sec-num-Chemistry}, 
and investigate different SF and SN feedback models, as summarized below. 

\begin{itemize} 
\item {\it E35nw}   % EffMod25-NW 
- Effective model (\S\ref{sec-num-EffMod}), no wind, no kinetic SN feedback, run NW of \citet{Barai13}. 

\item {\it E35rvw}   % EffMod25-RVW 
- Effective model, radially varying wind of velocity $v_w(r)$ (Equation \ref{eq-vSteidel}) 
with fixed parameter values described in \S\ref{sec-num-EffMod}, 
its speed going up to $v_{max} = 800$ km/s, run RVWa of \citet{Barai13}. 

\item {\it E25cw}   % EffMod18 
- Effective model, energy-driven wind with constant velocity $v_w = 350$ km/s. 

\item {\it M25std}   % Muppi18 
- Muppi model (\S\ref{sec-num-Muppi}) of SF and SN feedback, ``standard'' parameter values: 
$f_{\rm fb,out} = 0.2$, $f_{\rm fb,kin} = 0.6$, $P_{\rm kin} = 0.03$. 

\item {\it M25a}   % Muppi18-Fth04fk04 
- Muppi model, higher thermal and lower kinetic feedback energy fraction: 
$f_{\rm fb,out} = 0.4$, $f_{\rm fb,kin} = 0.4$, $P_{\rm kin} = 0.03$. 

\item {\it M25b}   % Muppi18-Fth02fk08 
- Muppi model, higher kinetic feedback energy fraction: 
$f_{\rm fb,out} = 0.2$, $f_{\rm fb,kin} = 0.8$, $P_{\rm kin} = 0.03$. 

\item {\it M25c}   % Muppi18-onepercent 
- Muppi model, less-efficient kinetic feedback, lower probability of kicking gas particles into wind: 
$f_{\rm fb,out} = 0.2$, $f_{\rm fb,kin} = 0.6$, $P_{\rm kin} = 0.01$. 
% kicking a smaller number of particles. 

\item {\it M25d}   % Muppi18-sixpercent 
- Muppi model, more-efficient kinetic feedback, higher probability of kicking gas particles into wind: 
$f_{\rm fb,out} = 0.2$, $f_{\rm fb,kin} = 0.6$, $P_{\rm kin} = 0.06$. 

\item {\it M50std}   % Muppi36 
- Muppi model, larger box, standard parameter values: 
$f_{\rm fb,out} = 0.2$, $f_{\rm fb,kin} = 0.5$, $P_{\rm kin} = 0.03$. 

\end{itemize} 

We compute the total gas metallicity, $Z_{\rm gas}$, 
as the ratio of all metal mass to the total gas mass for each gas particle. 
Abundance ratios are expressed in terms of the Solar metallicity, 
which is $Z_{\odot} = 0.0122$ (mass fraction of all metals in Sun) 
derived from the compilation by \citet{Asplund05}.

\subsection{Outflow Measurement Technique} 
\label{sec-num-FlowMeasureTech} 

We measure outflow of a galaxy by tracking the high-speed gas particles belonging to it. 
We identify halos in the simulations using the {\it Friends-of-Friends} (FOF) group finder algorithm, 
which gives the virial radius $R_{\rm vir}$. 
Subsequently we track galaxies using the subhalo finder {\it SubFind}, 
which identifies associated substructures to FOF halos. 
The center of each galaxy is considered as the location of the 
gravitational potential minimum of its subhalo. 
A minimum stellar mass of $10^{9} M_{\odot}$ is used to select the galaxy population, 
corresponding to $750$ star particles in the MUPPI model. 
Only the central galaxies are considered (not the satellites) for the outflow analysis. 
We define the {\it galaxy radius} of the centrals to be $R_{\rm gal} = R_{\rm vir} / 10$. 
Each central galaxy having $M_{\star} > 10^{9} M_{\odot}$ is post-processed as described below. 
First the coordinates are transformed such that the cold gas disk of the galaxy 
is rotating in the $[X-Y]$ plane, and $Z$-axis is the perpendicular direction. 

% $v_{\rm esc}=\sqrt{GM(r)/r}$ 

We aim to quantify the motion of gas particles caused by SN feedback, that are 
expected to move at speeds in excess of $\sim 300$ km/s in all models with kinetic feedback. 
Gas particles, in the form of hot gas or cold streams, 
will move within a DM halo with speeds of the order of the halo circular velocity, 
and typically below its escape velocity 
$v_{\rm esc} = \sqrt{2 G M_{\rm halo} / R_{\rm vir}}$ (for a galaxy halo mass $M_{\rm halo}$); 
we will call these ``virial'' motions. 
In order to quantify outflows one could estimate the mass outflow and inflow rate of all gas particles, 
and take their difference; 
this would single out SN-driven outflows as long as virial motions do not produce net changes in mass. 
This assumption may be correct in the inner parts of relatively small and slowly-evolving DM halos, 
but is surely incorrect when applied at the virial radius; 
in this case we would measure the net effect of outflows and cosmological infall. 

Another option is to select only gas particles 
whose radial velocity is positive and exceeds $v_{\rm esc}$. 
This option would have the merit of selecting outflowing gas particles 
that can make their way out of the halo, 
if hydrodynamical interactions with the halo gas do not slow them significantly. 
In less-massive halos, where $v_{\rm esc}$ is much smaller 
than the typical velocity of gas particles subject to kinetic feedback, 
such particles will clearly separate out in a radial distance--radial velocity plot 
(e.g., Fig.~2 of \citealt{Barai13}). 
In more-massive halos, however, 
outflowing particles will have similar speeds as those subject to virial motion, 
and the distinction will be more difficult. 
In this case, a selection based on the escape velocity will have the merit to measure 
not the whole outflow, but the fraction that can truly escape the halo. 

Any measure of average speed of outflowing particles, $\langle v_{\rm out} \rangle$, 
will be affected by the assumed lower velocity threshold. 
As a consequence, we expect $\langle v_{\rm out} \rangle$ to scale with halo mass 
when $v_{\rm esc}$ is used as a velocity threshold, 
even if kinetic feedback produced exactly the same velocities in all galaxies. 
Therefore, to study the relation of the average outflow velocity with respect to galaxy properties, 
we will use a fixed velocity threshold, 
and exclude all the halos with escape velocity higher than that threshold 
since the measure there would be significantly affected by virial motions. 
A value of $300$ km/s is a good compromise that allows to measure outflows 
for many galaxies excluding only a few, most-massive halos. 
However, the quantification of the mass loading will be preferentially done 
using the escape velocity threshold. 

% The outflow we want to quantify is SN feedback-driven motion of the gas moving outward in a galaxy, 
% as opposed to e.g.~one driven by thermal velocity giving rise to essentially random motion. 
% Hence we must define an appropriate lower limit to the velocity of the gas being counted, 
% so that the feedback driven motion is measured. 
% $v_{\rm esc} = \sqrt{2 G M_{\rm halo} / R_{\rm vir}}$ for a galaxy halo mass $M_{\rm halo}$  

We also need to define a volume around a galaxy where to measure the outflow, 
a region most expected to contain the gas ejected by SN feedback. 
The MUPPI sub-resolution model deposits thermal and kinetic energy from SN 
to neighboring gas inside a cone with semi-aperture angle $\theta = 60^{\circ}$, 
therefore to detect outflows we use two cylindrical volumes located above and below the galaxy 
and aligned with the galaxy disk. 
We use such cylindrical volumes to detect outflows in order to intercept the largest number 
of gas particles which have received SN feedback energy, and hence obtain a robust measurement. 
In addition, to better assess and quantify the geometry of the outflow, 
we also measure outgoing gas around a galaxy using a spherical shell around it. 
Most of our results in \S\ref{sec-res-Flow} utilize the bi-cylinder approach, 
except \S\ref{sec-res-Flow-Mdot-Rvir-vs-Rgal} where we apply the spherical shell technique. 

Two cylindrical volumes are constructed, 
at a distance of $R_{\rm gal}$ above and below the galaxy disk plane, along the $Z$-direction. 
Each cylinder has a height of $h_{\rm cyl} = R_{\rm gal}$. 
The cylinder radius extends beyond $R_{\rm gal}$, and the excess length subtends an angle of $60^{\circ}$ 
with a plane perpendicular to the disk at $R_{\rm gal}$. 
Mathematically, the cylinder radius is thus $(R_{\rm gal} + R_{\rm gal} \tan 60)$. 
The $60^{\circ}$ angle is chosen to coincide with the opening angle 
within which gas particles are kicked during SN kinetic feedback of MUPPI sub-resolution model. 

Lets denote that the $i$'th gas particle has a mass $m_i$, z-velocity $v_{z, i}$, 
and lies at a z-coordinate of $z_i$. 
All the gas particles are selected, whose positions lie inside either cylinder, 
and moving at a high-speed such that: $|v_{z, i}| > v_{\rm lim,out}$, 
or the z-magnitude of velocity is higher than a limiting speed. 
Here $v_{\rm lim,out}$ can be a fixed value, or the escape velocity. 
The results in \S\ref{sec-res-Flow} are illustrated using both the limits for a few runs, 
and the final physical choices are described next. 
If the z-velocity is directed outward $(z_i \cdot v_{z, i} > 0)$ 
then the gas particle is counted as outflow. 

The outflow velocity, $v_{\rm out}$, is estimated from $|v_{z}|$. 
One of our goals is to infer a physical galaxy property (e.g.~mass, SFR) with which 
$v_{\rm out}$ correlates better. 
Selecting gas particles above $v_{\rm esc}$ brings in 
some inherent correlations between the resulting computed $v_{\rm out}$ and galaxy mass, 
because of the dependence of $v_{\rm esc}$ on $M_{\rm halo}$ by definition. 
Such built-in relations are unwanted and must be eliminated. 
Therefore, while calculating $v_{\rm out}$, we choose a constant value as the limiting speed 
for selecting gas particles: $v_{\rm lim,out} = 300$ km/s. 
The mass-weighted average outflow velocity is determined as: 
\begin{equation} 
\label{eq-FlowProp-MassAvg} 
\langle v_{\rm out} \rangle = \frac{\sum_{i=1}^{n_{\rm out}(|v_{z, i}| > 300)} |v_{z, i}| m_i} {\sum_{i=1}^{n_{\rm out}(|v_{z, i}| > 300)} m_i}. 
\end{equation} 
The limiting halo mass for which $v_{\rm esc} = 300$ km/s is, 
$M_{\rm halo, lim} = 1.3 \times 10^{12} M_{\odot}$. 
Measuring $v_{\rm out}$ by selecting gas above $300$ km/s ensures that 
outflows in galaxies less massive than $M_{\rm halo, lim}$ can escape the halo potential, 
and might not escape away in galaxies more massive than $M_{\rm halo, lim}$. 
This is accordingly noted in our analysis presented in \S\ref{sec-res-Flow-Vel}. 

On the other hand, the mass outflow rate, $\dot{M}_{\rm out}$, should be evaluated over 
all the gas which can escape the galaxy total gravitational potential. 
Therefore, $v_{\rm lim,out} = v_{\rm esc}$ is used here. 
We calculate $\dot{M}_{\rm out}$ by summing over all the $n_{\rm out}$ 
gas particles having $|v_{z, i}| > v_{\rm esc}$ : 
\begin{equation} 
\label{eq-FlowProp-Mdot} 
\dot{M}_{\rm out} = \sum_{i=1}^{n_{\rm out}(|v_{z, i}| > v_{\rm esc})} \frac{m_i |v_{z, i}|} {h_{\rm cyl}}. 
\end{equation} 

We use a sphere technique for measuring outflows in \S\ref{sec-res-Flow-Mdot-Rvir-vs-Rgal}. 
{\it A spherical shell} volume of thickness $h_{\rm sph}$ is constructed around each galaxy center. 
The gas particles lying inside the spherical shell, and moving 
radially faster than the escape velocity at that radius: $|v_r| > v_{\rm esc}(r)$, are selected. 
If the radial velocity is directed outward $(v_r > 0)$ then the gas particle is counted as outflow. 
We calculate the mass outflow rate as: 
\begin{equation} 
\label{eq-Mdot-Sph} 
\dot{M}_{\rm out} = \sum_{i=1}^{n_{\rm out}} \frac{m_i |v_{r, i}|} {h_{\rm sph}}. 
\end{equation} 
For $r = R_{\rm gal}$, the inner radius of the shell is located 
at a distance of $R_{\rm gal}$ from galaxy center, $h_{\rm sph} = R_{\rm gal}$, 
and $|v_r| > v_{\rm esc}(R_{\rm gal})$ is used. 
For $r = R_{\rm vir}$, 
the shell inner radius is at a distance $0.9 R_{\rm vir}$, $h_{\rm sph} = 0.1 R_{\rm vir}$. 
and $|v_r| > v_{\rm esc}(R_{\rm vir})$ is used. 

% The halo mass ($M_{\rm halo}$) and virial radius in comoving coordinates ($R_{200}$) are related 
% such that $R_{200}$ encloses a density $200$ times the mean comoving density of the Universe: 
% \begin{equation} 
% \label{eq-Mhalo} 
% M_{\rm halo} = \frac{4 \pi}{3} R_{200}^3 \left(200 \rho_{\rm crit} \Omega_{M,0}\right). 
% \end{equation} 
% Here $\rho_{\rm crit} = 3 H_0^2 / (8 \pi G)$ is the present critical density. 

% modifying the methodology developed in \citet{Ragagnin13}. 
% We estimate various physical properties of the outflowing gas, thus $\langle x \rangle$ can be: 
% density $(\rho_{\rm gas})$, temperature $(T_{\rm gas})$, metallicity $(Z_{\rm gas})$. 
% The process is repeated for all the selected galaxies, and results are presented in \S\ref{sec-res-Flow}. 
% and $n_{\rm in}$ % and inflow, respectively. 
% \langle x_{\rm in}  \rangle = \frac{\sum_{i=1}^{n_{\rm in}}  x_i m_i} {\sum_{i=1}^{n_{\rm in}}  m_i}. 
% \dot{M}_{\rm in}  = \sum_{i=1}^{n_{\rm in}}  \frac{m_i |v_{z, i}|} {h_{\rm cyl}}. 
% Thus the $v_{\rm out}$ measured using our methodology 
% we consider only the lower-mass ($M_{\rm halo} < M_{\rm halo, lim}$) galaxies. 

\section{Results and Discussion} 
\label{sec-results}

\subsection{Star Formation Rate Density} 
\label{sec-res-SFRD} 

%%%%%%%%%%%%%%%%%%%%%%%%%%%%%%%%%%%%%%%%%%%%%%%%%%%%%%%%%%%%%%%%%%%%%%%% 
% FIGURE 1 
\begin{figure*} 
\centering 
\includegraphics[width = 0.7 \linewidth]{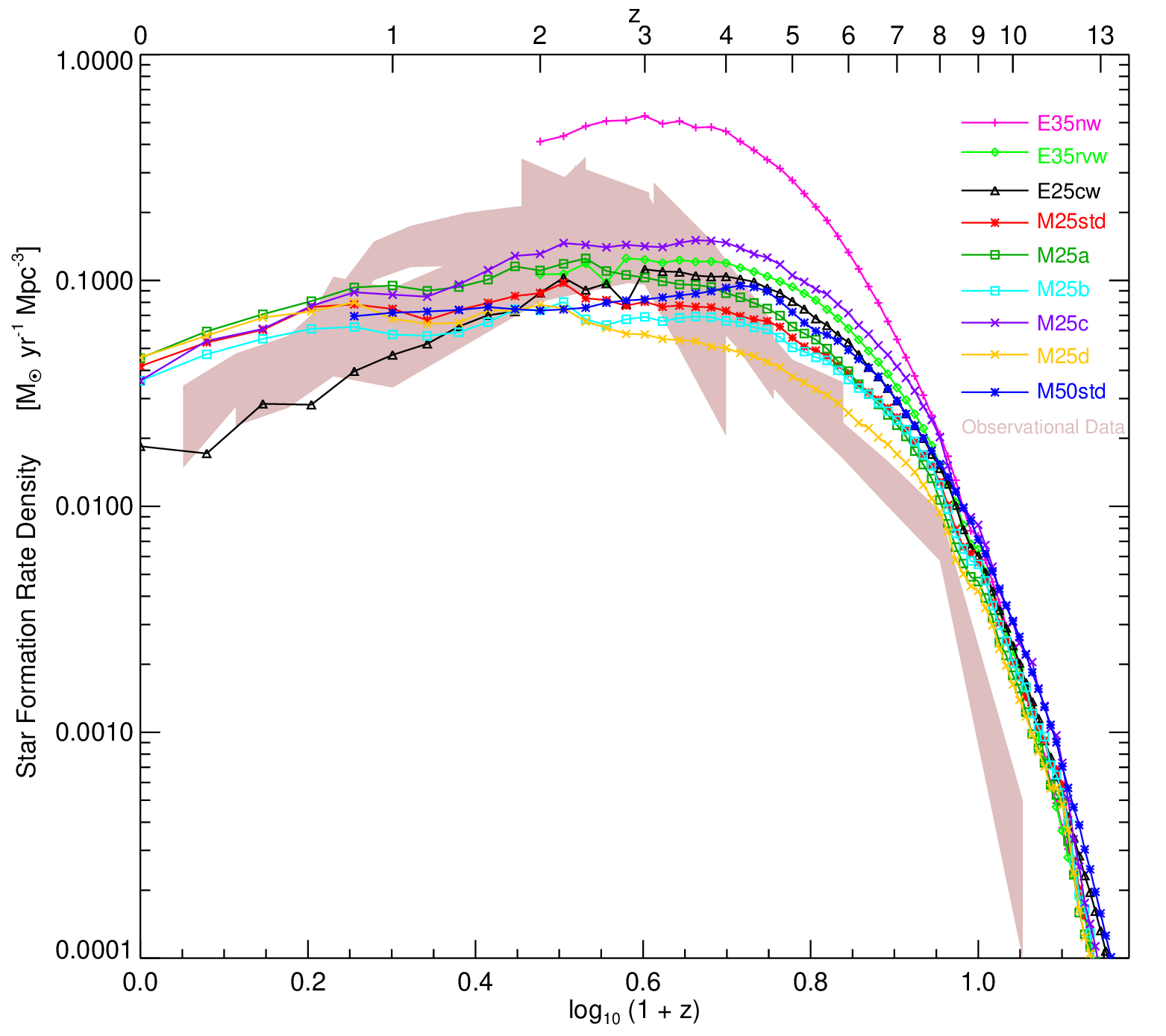} 
\caption{ 
Total star formation rate density in simulation volume as a function of redshift, 
with the different models labelled by the colour and plotting symbol. 
The grey shaded region denote a combination of observational data range from \citet{Cucciati12}, 
and the compilations therein originally from \citet{PerezGonzalez05}, \citet{Schiminovich05}, 
\citet{Bouwens09}, \citet{Reddy09}, \citet{Rodighiero10}, \citet{vanderBurg10}, \citet{Bouwens12}. 
The comparison is described in \S\ref{sec-res-SFRD}. 
} 
\label{fig-SFRD} 
\end{figure*} 
% \citet{Steidel99}, % \citet{Ouchi04}, 
%%%%%%%%%%%%%%%%%%%%%%%%%%%%%%%%%%%%%%%%%%%%%%%%%%%%%%%%%%%%%%%%%%%%%%%% 

The global star formation rate density (SFRD) as a function of redshift is plotted in Fig.~\ref{fig-SFRD}, 
with the nine runs labelled by the different colours and plotting symbols. 
The SFRD (in $M_{\odot}$ yr$^{-1}$ Mpc$^{-3}$) is computed by summing over 
all the SF occurring in each simulation box at a time, 
and dividing it by the time-step interval and the box volume. 
Observational data ranges are shown as the grey shaded region, 
taken mainly from \citet{Cucciati12}, and the compilations therein originally from 
\citet{PerezGonzalez05}, \citet{Schiminovich05}, 
\citet{Bouwens09}, \citet{Reddy09}, \citet{Rodighiero10}, \citet{vanderBurg10}, \citet{Bouwens12}. 

SN feedback clearly has a significant impact on the SFRD; 
compared to {\it E35nw}, SF is reduced by a factor of several in the other runs at $z < 8$. 
The stronger kinetic feedback runs ({\it M25b} and {\it M25d}) quench SFR more than the others. 
In fact, case {\it M25d} produce the lowest SFRD from high-$z$ up to $z = 2$, 
but at $z < 2$ run {\it E25cw} has the lowest SFRD. 
Among the feedback runs, {\it M25c} produces the highest SFRD; 
implying that a $1 \%$ probability of kicking gas particles into wind 
yields the least efficient kinetic feedback. 

The scatter in the SFRD values within the MUPPI runs are larger at $z > 1$, and reduces at $z < 1$ 
due to overall gas depletion. 
The SFRD is higher in {\it M50std} than in {\it M25std} from high-$z$ up to $z \sim 3$, 
and becomes comparable later. 
This is because, despite having the same parameter values and resolution, runs {\it M50std} and {\it M25std} 
uses different cooling functions (\S\ref{sec-num-cool}). 

At earlier cosmic epochs $z > 5$, the SFRD in the simulations 
is systematically higher, reaching $2 - 5$ times the observed values. 
Later at $z < 5$, 
most of the models overall lie within the ranges of SFRD produced by the different observations. 
However at $z < 0.7$, 
the MUPPI models produce a higher and the Effective model a lower SFRD than the observations. 
The low-$z$ discrepancy is because we do not have AGN feedback, 
which when present is expected to quench additional SF, bringing the MUPPI models closer to observations. 
Studying the shape evolution, 
most of the simulations reach a maximum SFRD in the form of a plateau between $z = 2 - 4$, 
and the SFRD decreases steeply at $z > 4$ and gradually at $z < 2$. 
This is contrary to the peak of observed SFRD at $z \sim 2$. 

We compare our results with two contemporary cosmological hydrodynamical simulations. 
At $z \geq 3$ the SFRD in our simulations is higher than that 
obtained in $Illustris$ \citep{Vogelsberger14}, 
our SFRD exhibits a plateau maximum and not a well-defined peak at $z \sim 2$ like them, 
and at $z = 0$ also our SFRD is higher. 
Our results are more akin to that of some runs in $OWLS$ \citep{Schaye10} 
which show a SFRD plateau between $z = 2 - 4$. 

% the SFRD in simulations is systematically lower than observations between $z \sim 1 - 3$; 
% The contribution of extra 3 elements in run {\it M25std} decreases the gas cooling rate 
% at high-$z$, causing a lower SFRD. 
% 8 elements are tracked in run {\it M50std} % uses cooling tables of \citet{Sutherland93} , 
% while 11 elements in {\it M25std},  % uses tables of \citet{Wiersma09a} which tracks 
% We see that no simulation model can fit the data from observations at all redshifts. 
% relevant box-size; halos start forming in box {\it M50std} earlier than in {\it M25std}, 
% as well as are more-massive and more-numerous. 
% From $z = 10$, {\it M18a} and {\it M25d} has the lowest SFR, 
% while {\it M25c} and {\it M50std} has the highest (2 times higher) SFR. 
% Effects of box-size are visible in our results. 
% Run CW of the LB series is the model providing the best-fit to the observations at lower-$z$. 
% Each data set is shown with a different plotting symbol as listed next. 
% Taken collectively, there is a better match of simulations with observations at low-$z$. 

\subsection{Single Galaxy}    % and Outflow Measurement Volume 
\label{sec-res-SingGal} 

%%%%%%%%%%%%%%%%%%%%%%%%%%%%%%%%%%%%%%%%%%%%%%%%%%%%%%%%%%%%%%%%%%%%%%%% 
% FIGURE 2 
\begin{figure*} 
\centering 
\includegraphics[width = 1.05 \linewidth]{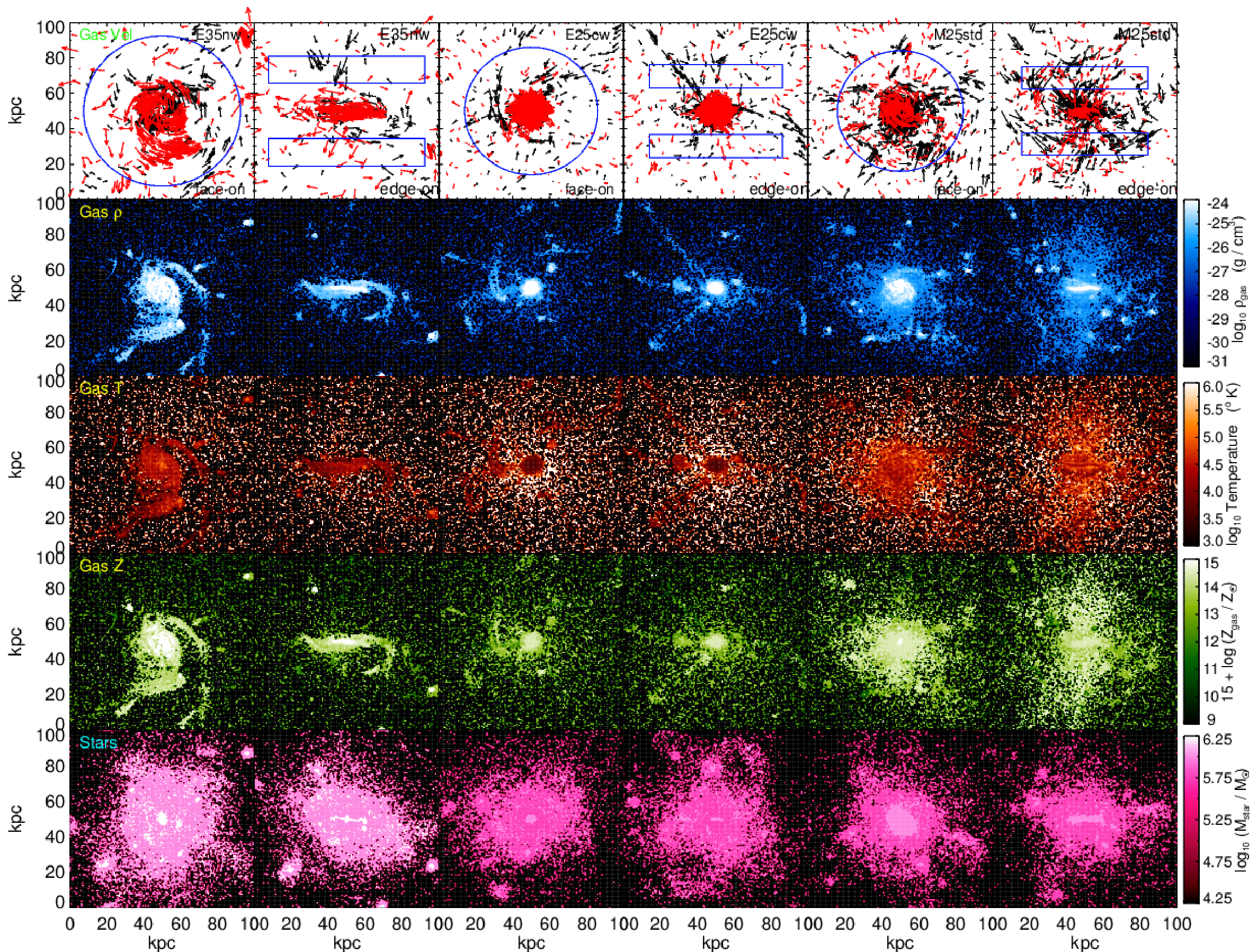} 
\caption{ 
Projection of gas kinematics (top four rows) and stars (bottom row) in the face-on (left) 
and edge-on (right) planes of a $(100$ kpc$)^3$ volume 
centered around a massive galaxy at $z = 2$, in runs: 
{\it E35nw} (left two columns), {\it E25cw} (middle two columns), and {\it M25std} (right two columns). 
First row depicts the velocity vectors of $7 \%$ of all the gas particles inside the plotted volume, 
with the outflowing ($v_r > 0$) particles denoted as red, and the inflowing ($v_r < 0$) as black. 
Second row shows gas density, third row is gas temperature, and fourth row is gas total metallicity. 
Bottom row presents the projected stellar mass. 
In the top row, the blue circle in the face-on panels and the double blue rectangles 
in the edge-on panels illustrate the projected bi-cylinder volume used to measure outflows. 
} 
\label{fig-OneGal-Pos-Vel-rho-Z} 
\end{figure*} 
%%%%%%%%%%%%%%%%%%%%%%%%%%%%%%%%%%%%%%%%%%%%%%%%%%%%%%%%%%%%%%%%%%%%%%%% 

The gas distribution of a massive galaxy of total mass 
$M_{\rm halo} = 3.4 \times 10^{12} M_{\odot}$ (left two columns) and 
$1.8 \times 10^{12} M_{\odot}$ (remaining columns) at $z = 2$ is plotted in 
Fig.~\ref{fig-OneGal-Pos-Vel-rho-Z} for three runs. 
Each of the five rows shows a gas/star property, 
projected in the face-on (left) and edge-on (right) planes of a $(100$ kpc$)^3$ volume. 
First row depicts the velocity vectors of $7 \%$ gas particles, 
with the outgoing ($v_r > 0$) particles denoted as red, and incoming ($v_r < 0$) gas as black. 
Here the blue circle in the face-on panels and the double blue rectangles 
in the edge-on panels illustrate the projected bi-cylinder volume used to measure the outflow 
(\S\ref{sec-num-FlowMeasureTech}) of each galaxy. 
Gas density is in the second row, temperature in the third, 
and total metallicity is plotted in the fourth row. 

{\it E35nw} and {\it M25std} present the formation of a gas disk with 
extended spiral arms and tidal features. 
In the no-wind case {\it E35nw}, the gas disk is bigger in size, more massive, more metal enriched; 
and there is no prominent outflow. 
While in run {\it E25cw} the central gas distribution is spheroidal, 
and most of the outflowing gas lies inside $r < 20$ kpc, because here 
the wind kick velocity $v_w = 350$ km/s is too small to drive large-scale outflows. 
The Muppi run {\it M25std} produces a well-developed gas outflow 
propagating perpendicular to the galaxy disk, escaping to $r > 30$ kpc from the galaxy center. 
Metals are more distributed in runs {\it E25cw} and {\it M25std}, 
since SN winds carry the metals out from the SF regions and enrich the CGM. 
The fourth row also shows that the {\it M25std} outflow (right two panels) is more metal enriched 
at $30$ kpc from the galaxy, than the {\it E25cw} case (middle two panels). 
Stellar mass in the bottom row reveals a central disk-like structure, surrounded by a larger stellar halo. 
The stellar disk is thinner in the no-wind case {\it E35nw}, than in the MUPPI run {\it M25std}. 

% the strong gravitational potential of the halo causes the gas to fall back inward. 

\subsection{Outflow Properties of Galaxies at $z = 2$} 
\label{sec-res-Flow}

We measure the gas outflow properties of galaxies 
using the bi-cylinder technique described in \S\ref{sec-num-FlowMeasureTech}, 
with a limiting velocity both fixed and scaled with $v_{\rm esc}$, 
by analysing central galaxies having $M_{\star} > 10^{9} M_{\odot}$. 
The results of all the nine simulations at $z = 2$ are presented in this section. 

We study the trends of outflow $v_{\rm out}$ and $\dot{M}_{\rm out}$ with basic galaxy properties: 
(i) halo mass, 
(ii) gas mass, 
(iii) stellar mass, 
(iv) SFR (sum of all the instantaneous star formation occurring in the gas belonging to a galaxy), 
(v) gas consumption timescale = gas mass / SFR, 
(vi) star formation timescale = stellar mass / SFR; 
attempting to find possible correlations. 
We find positive correlations of the outflow quantities with galaxy mass and SFR. 
However the correlation with galaxy SFR is the tightest, hence we present only these here.

\subsubsection{Outflow Velocity} 
\label{sec-res-Flow-Vel} 

%%%%%%%%%%%%%%%%%%%%%%%%%%%%%%%%%%%%%%%%%%%%%%%%%%%%%%%%%%%%%%%%%%%%%%%% 
% FIGURE 3 
\begin{figure*} 
\centering 
\includegraphics[width = 0.9 \linewidth]{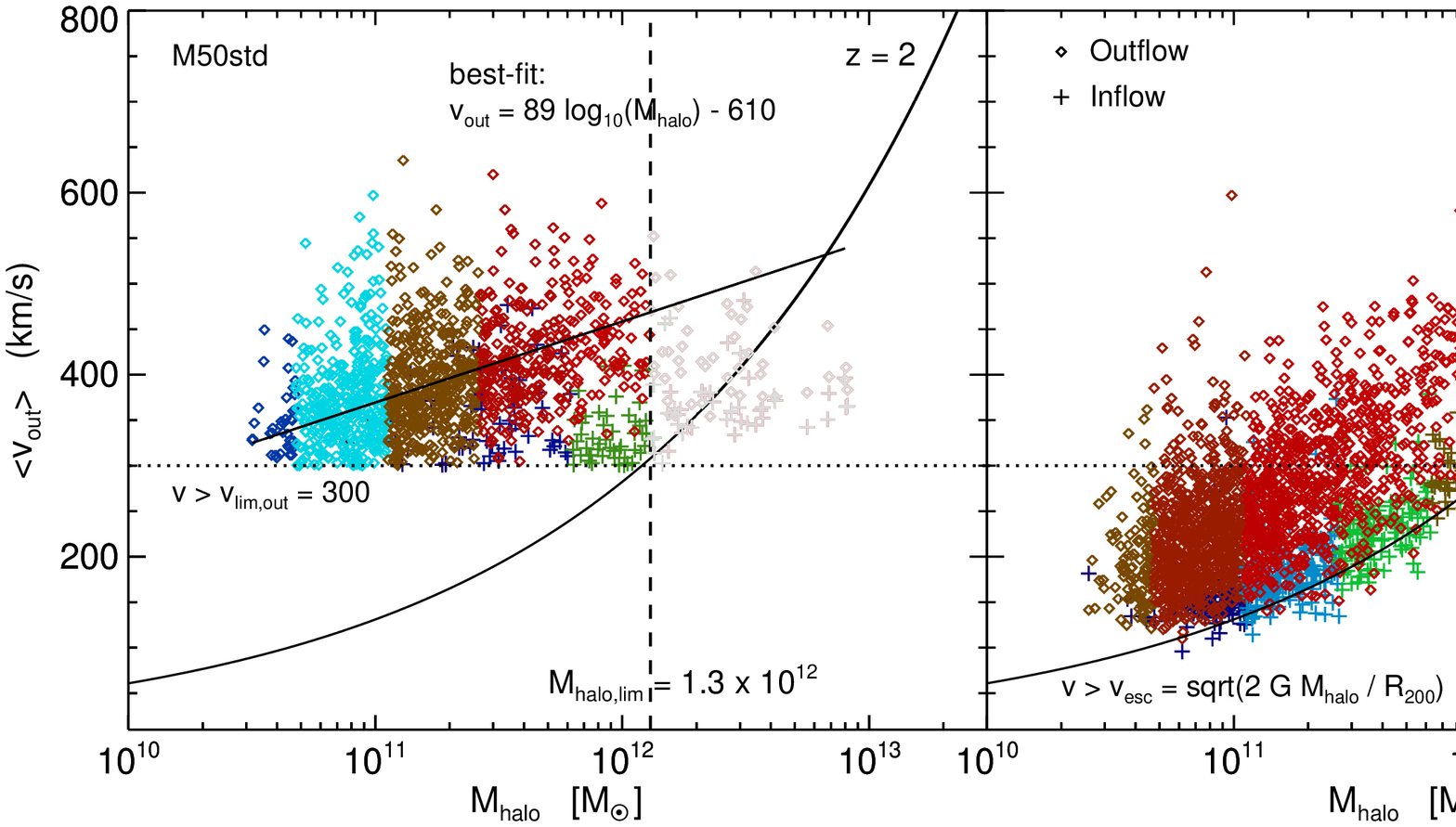} 
\caption{ 
Outflow velocity mass-weighted average, $\langle v_{\rm out} \rangle$, 
as a function of halo mass of galaxies in run {\it M50std}. 
The two panels show different methods of computing $v_{\rm out}$, by selecting gas particles above: 
a constant limiting speed $|v_{z, i}| > 300$ km/s (horizontal {\it black dotted} line) in the left 
(Eq.~\ref{eq-FlowProp-MassAvg} in \S\ref{sec-num-FlowMeasureTech}), 
and the escape velocity $|v_{z, i}| > v_{\rm esc}$ ({\it black solid} curve) at the right. 
Each point is one galaxy, and the plotting symbol indicates the flow direction of the measurements: 
outflows are denoted by {\it diamonds}, inflows by {\it plus symbols}. 
The plotting colour depicts the number fraction of galaxies where outflow or inflow is detected 
in bins of halo mass. 
The grey points in the left panel (includes both outflow and inflow) mark galaxies more massive 
than $M_{\rm halo, lim} = 1.3 \times 10^{12} M_{\odot}$ (vertical {\it black dashed} line), 
where the outflows might not escape the halo potential. Details in \S\ref{sec-res-Flow-Vel}. 
} 
\label{fig-Vout-vs-MFoF} 
\end{figure*} 
%%%%%%%%%%%%%%%%%%%%%%%%%%%%%%%%%%%%%%%%%%%%%%%%%%%%%%%%%%%%%%%%%%%%%%%% 

The velocity of outflow, $\langle v_{\rm out} \rangle$, as a function of halo mass of galaxies 
in run {\it M50std} is plotted in Fig.~\ref{fig-Vout-vs-MFoF}, 
illustrating the two different methods of computing $v_{\rm out}$. 
Outflowing gas particles are selected when their $z$-velocity component is above: 
(i) a constant limiting speed or $|v_{z, i}| > 300$ km/s (Eq.~\ref{eq-FlowProp-MassAvg}) 
in the left panel, and 
(ii) the escape velocity or $|v_{z, i}| > v_{\rm esc}$ at the right, 
as described in \S\ref{sec-num-FlowMeasureTech}. 
As a test we present the inflows: gas particles having velocities over the threshold 
and $z_i \cdot v_{z, i} < 0$, as well in this figure. 
The flow direction is indicated by the plotting symbols: 
outflows as {\it diamonds}, and inflows as {\it plus symbols}. 
The plotting colour depicts the number fraction of galaxies where outflow or inflow is detected 
in bins of halo mass. 
The grey points in the left panel (includes both outflow and inflow) mark galaxies more massive 
than $M_{\rm halo, lim} = 1.3 \times 10^{12} M_{\odot}$ (vertical {\it black dashed} line). 
In the right panel the inflow velocities are spread very close to the $v_{\rm esc}$ curve, 
because inflows consist of thermal velocity and random motion of the gas, 
and by imposing a lower cutoff of $v_{\rm esc}$ we measure the motion occurring close to $v_{\rm esc}$. 
Whereas most the outflow velocities are larger (by a few times) than $v_{\rm esc}$ at all halo masses; 
which demonstrates that we are measuring non-thermal gas outflow 
driven by stellar and SN feedback processes. 
The left panel, by selecting gas above a fixed cutoff of $300$ km/s, shows a positive correlation 
of the measured outflow velocity with halo mass. 

%%%%%%%%%%%%%%%%%%%%%%%%%%%%%%%%%%%%%%%%%%%%%%%%%%%%%%%%%%%%%%%%%%%%%%%% 
% FIGURE 4 
\begin{figure*} 
\centering 
\includegraphics[width = 0.9 \linewidth]{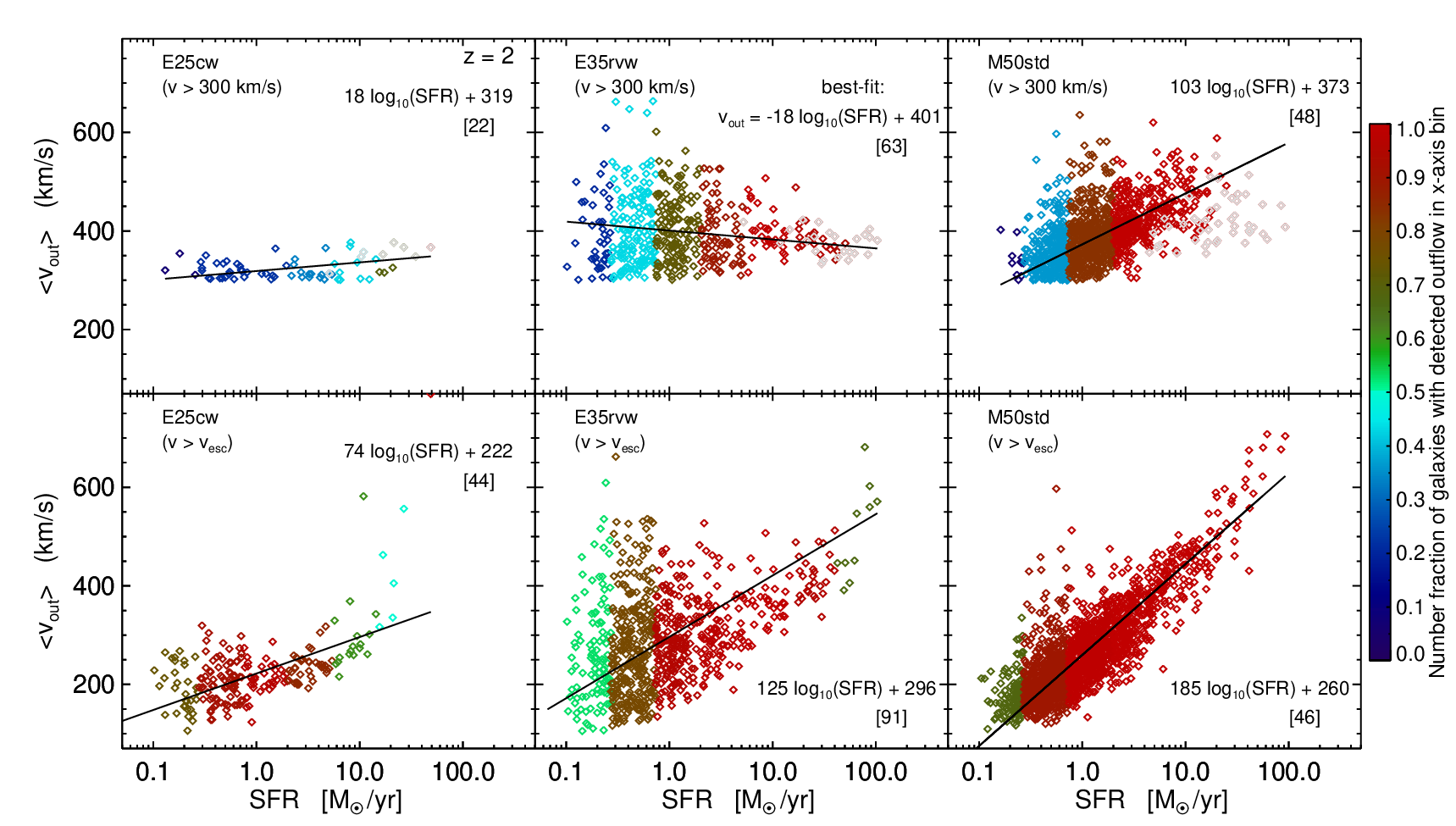} 
\caption{ 
Outflow velocity as a function of SFR of galaxies in three runs: 
{\it E25cw} in the left column, {\it E35rvw} at the middle, and {\it M50std} in the right. 
The two rows show different methods of computing outflow, by selecting gas particles above: 
a constant limiting speed $|v_{z, i}| > 300$ km/s at the top, 
and the escape velocity $|v_{z, i}| > v_{\rm esc}$ in the bottom. 
The plotting colour depicts the number fraction of galaxies where outflow is detected in bins of galaxy SFR. 
The grey points in the top row mark galaxies more massive than 
$M_{\rm halo, lim} = 1.3 \times 10^{12} M_{\odot}$, where the outflows might not escape the halo potential. 
The black line is the best-fit relation between $v_{\rm out}$ and $\log_{10}{\rm(SFR)}$ 
obtained by an outlier-resistant two-variable linear regression. 
The fit coefficients and the ``standard deviation'' of the fit's residuals (within square brackets) 
are written inside the panels, using units same as the plotted axes. 
} 
\label{fig-Vout-vs-SFR-3runs-2cutoff} 
\end{figure*} 
%%%%%%%%%%%%%%%%%%%%%%%%%%%%%%%%%%%%%%%%%%%%%%%%%%%%%%%%%%%%%%%%%%%%%%%% 

The outflow velocity as a function of SFR of galaxies 
is plotted in Fig.~\ref{fig-Vout-vs-SFR-3runs-2cutoff} for three runs, 
illustrating the two different methods of computing outflow in the two rows. 
% Gas particles are selected above: 
% (i) a constant limiting speed $|v_{z, i}| > 300$ km/s in the top row, and 
% (ii) the escape velocity $|v_{z, i}| > v_{\rm esc}$ at the bottom. 
The plotting colour depicts the number fraction of galaxies where outflow is detected in bins of SFR, 
as indicated by the colorbar on the right. 
The differences between the two rows is the largest for the Effective models 
{\it E25cw} and {\it E35rvw} (left two columns), 
because these runs have energy-driven wind where $v_w$ is not dependent on galaxy mass. 
{\it E25cw} has $v_w = 350$ km/s, 
and using a fixed lower cutoff of $300$ km/s selects only a few gas particles in the lower-mass galaxies. 
For the MUPPI run {\it M50std}, the two distinct lower cutoffs make a small difference. 

%%%%%%%%%%%%%%%%%%%%%%%%%%%%%%%%%%%%%%%%%%%%%%%%%%%%%%%%%%%%%%%%%%%%%%%% 
% FIGURE 5 
\begin{figure*} 
\centering 
\includegraphics[width = 1.0 \linewidth]{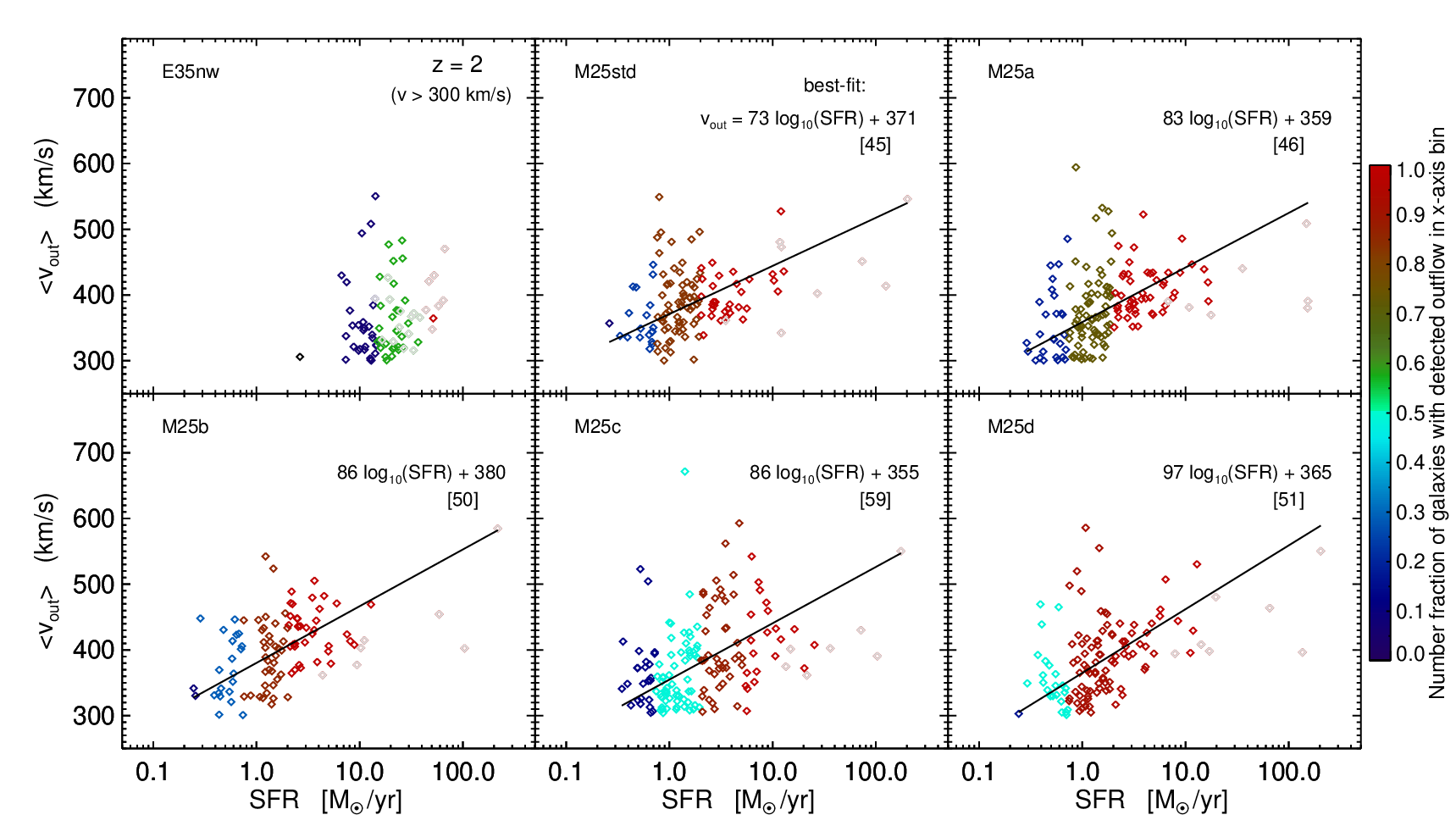} 
\caption{ 
Outflow velocity as a function of SFR of galaxies for six simulations, one in each panel. 
The format is the same as the top row of Fig.~\ref{fig-Vout-vs-SFR-3runs-2cutoff}. 
The plotting colour depicts the number fraction of galaxies where outflow is detected in bins of SFR. 
The grey points mark galaxies more massive than $M_{\rm halo, lim} = 1.3 \times 10^{12} M_{\odot}$, 
where the outflows might not escape the halo potential. 
The black line in the latter five panels is the best-fit relation 
between $v_{\rm out}$ and $\log_{10}{\rm(SFR)}$. 
} 
\label{fig-Vout-vs-SFR} 
\end{figure*} 
%%%%%%%%%%%%%%%%%%%%%%%%%%%%%%%%%%%%%%%%%%%%%%%%%%%%%%%%%%%%%%%%%%%%%%%% 

Fig.~\ref{fig-Vout-vs-SFR} presents the outflow velocity as a function of SFR of galaxies 
for the six remaining simulations, where outflow is measured 
by selecting gas above a fixed $|v_{z, i}| > 300$ km/s. 
The grey points mark galaxies more massive than $M_{\rm halo, lim} = 1.3 \times 10^{12} M_{\odot}$, 
where the outflows might not escape the halo potential. 

We see that kinetic SN feedback is needed to generate an outflow in the 
lower-SFR (consequently less-massive) galaxies. 
Therefore the no-wind run {\it E35nw} shows outflowing gas only in 
some galaxies with SFR $> 8 M_{\odot} /$ yr; 
this can be interpreted as the expected contamination from gravity-driven gas flows, and 
is roughly consistent with the statistics of inflows in the left panel of Fig.~\ref{fig-Vout-vs-MFoF}. 
In the other runs more than $50\%$ galaxies with SFR $\geq 1 M_{\odot} /$ yr have outflows. 
The number fraction of galaxies where outflow is detected 
(indicated by the plotting colour) increases with galaxy SFR. 

All the runs, except {\it E25cw}, present a large scatter. 
The Effective models produce, with low probability, an outflow velocity which shows no relation with SFR. 
Run {\it E25cw} has a constant $v_{\rm out} \sim 300$ km/s, 
since the input model is energy-driven wind with a fixed $v_w = 350$ km/s. 
Run {\it E35rvw} has a large scatter with $v_{\rm out} \sim 300 - 500$ km/s. 
It uses a radially varying wind of velocity $v_w(r)$ (Eq.~\ref{eq-vSteidel}) 
with the speed going up to $v_{max} = 800$ km/s at $R_{\rm eff} = 100 h^{-1}$ kpc. 
This turns out to be a strong wind for the low-SFR galaxies, where $v_{\rm out} \sim 650$ km/s is reached. 

The six MUPPI runs (five in Fig.~\ref{fig-Vout-vs-SFR} and one in Fig.~\ref{fig-Vout-vs-SFR-3runs-2cutoff}) 
display a positive correlation of $v_{\rm out}$ with galaxy SFR: 
$v_{\rm out}$ rising from $300$ km/s at low-SFR to $\sim 600$ km/s at high-SFR. 
It is the most steep in the larger-box run {\it M50std}, which has the highest number of galaxies. 
The positive correlation has the largest scatter in {\it M25c}, 
the least efficient kinetic feedback case with a $1 \%$ probability of kicking gas particles into wind.

\subsubsection{Mass Outflow Rate} 
\label{sec-res-Flow-Mdot} 

%%%%%%%%%%%%%%%%%%%%%%%%%%%%%%%%%%%%%%%%%%%%%%%%%%%%%%%%%%%%%%%%%%%%%%%% 
% FIGURE 6 
\begin{figure*} 
\centering 
\includegraphics[width = 0.9 \linewidth]{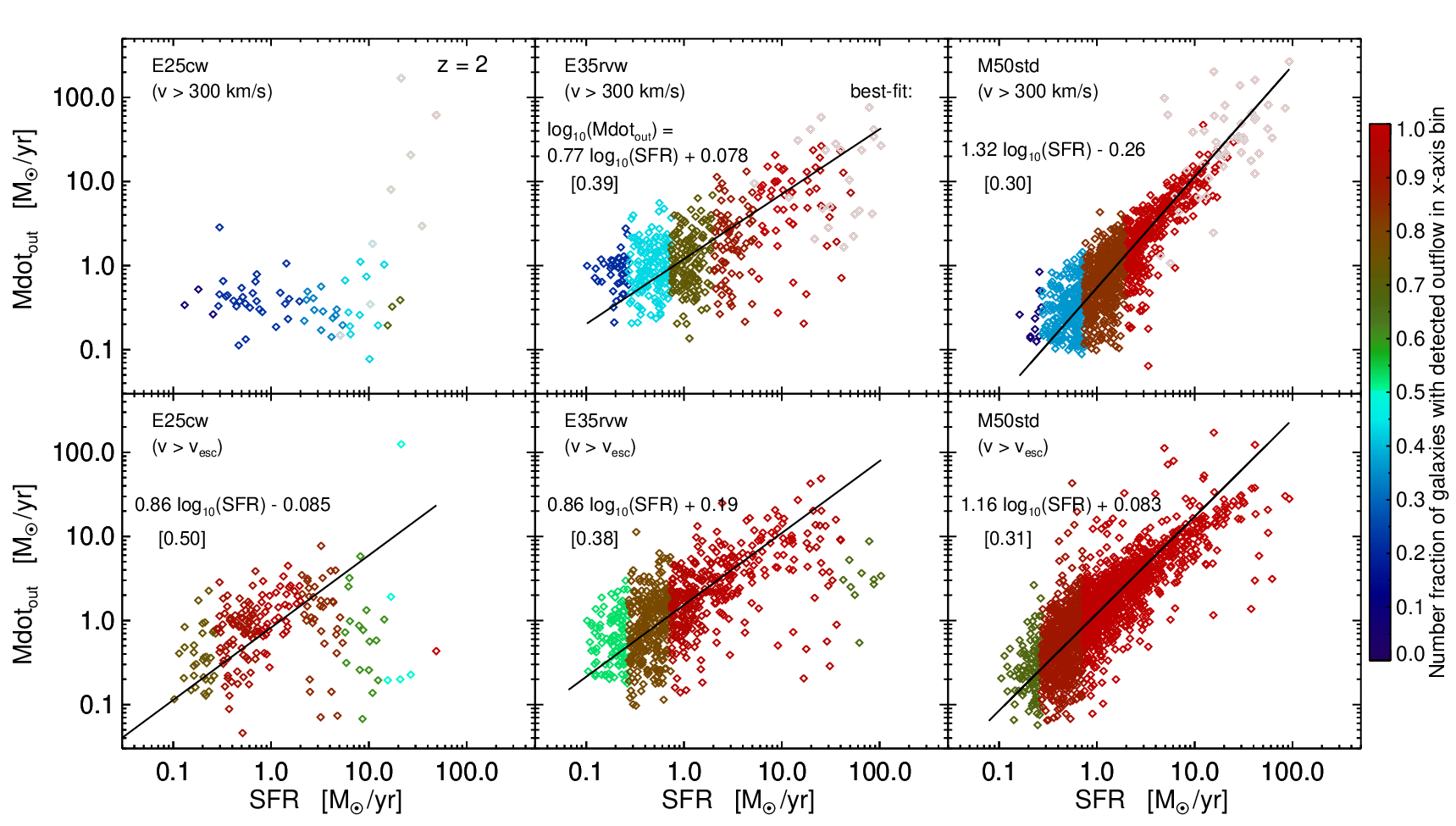} 
\caption{ 
Mass outflow rate, $\dot{M}_{\rm out}$ from Eq.~(\ref{eq-FlowProp-Mdot}), 
as a function of SFR of galaxies in three runs: 
{\it E25cw} in the left column, {\it E35rvw} at the middle, and {\it M50std} in the right. 
The two rows show different methods of computing outflow, by selecting gas particles above: 
a constant limiting speed $|v_{z, i}| > 300$ km/s at the top, 
and the escape velocity $|v_{z, i}| > v_{\rm esc}$ in the bottom. 
The plotting colour depicts the number fraction of galaxies where outflow is detected in bins of galaxy SFR. 
The grey points in the top row mark galaxies more massive than 
$M_{\rm halo, lim} = 1.3 \times 10^{12} M_{\odot}$, where the outflows might not escape the halo potential. 
The black line in the latter five panels is the best-fit relation between $\dot{M}_{\rm out}$ and SFR, 
with the coefficients of an outlier-resistant two-variable linear regression written inside the panels. 
Details in \S\ref{sec-res-Flow-Mdot}. 
} 
\label{fig-Mdot-vs-SFR-3runs-2cutoff} 
\end{figure*} 
%%%%%%%%%%%%%%%%%%%%%%%%%%%%%%%%%%%%%%%%%%%%%%%%%%%%%%%%%%%%%%%%%%%%%%%% 

The mass outflow rate, 
$\dot{M}_{\rm out}$ from Eq.~(\ref{eq-FlowProp-Mdot}) in \S\ref{sec-num-FlowMeasureTech}, 
as a function of SFR of galaxies, is plotted in Fig.~\ref{fig-Mdot-vs-SFR-3runs-2cutoff} for three runs, 
illustrating the two different methods of computing outflow in the two rows. 
Gas particles are selected above: 
(i) a constant limiting speed $|v_{z, i}| > 300$ km/s in the top row, and 
(ii) the escape velocity $|v_{z, i}| > v_{\rm esc}$ at the bottom. 
It is physically expected that gas moving faster than $v_{\rm esc}$ is able to escape the galaxy halo; 
therefore the bottom row gives 
a physically motivated estimate of the outflowing mass that can make its way out of the halo. 
The differences between the two rows is the largest for run {\it E25cw} in the left column; 
because it has energy-driven wind with $v_w = 350$ km/s, 
and using a fixed lower cutoff of $300$ km/s selects only a few gas particles in the lower-mass galaxies. 
For the other two runs with strong kinetic feedback: {\it E35rvw} and {\it M50std}, 
the two distinct lower cutoffs make very small difference. 

%%%%%%%%%%%%%%%%%%%%%%%%%%%%%%%%%%%%%%%%%%%%%%%%%%%%%%%%%%%%%%%%%%%%%%%% 
% FIGURE 7 
\begin{figure*} 
\centering 
\includegraphics[width = 0.9 \linewidth]{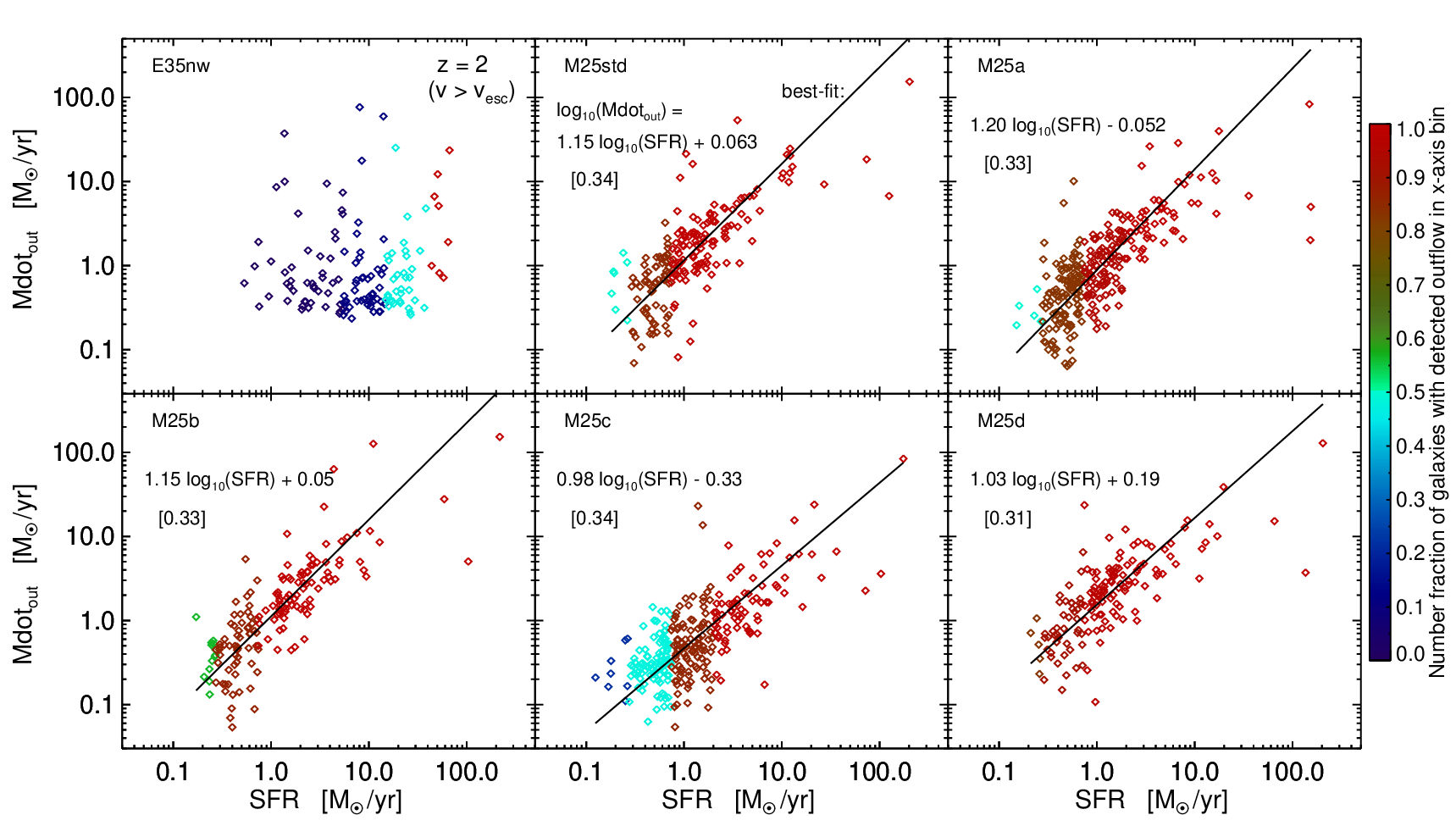} 
\caption{ 
Mass outflow rate as a function of SFR of galaxies, for six simulations one in each panel. 
The format is the same as the bottom row of Fig.~\ref{fig-Mdot-vs-SFR-3runs-2cutoff}. 
The plotting colour depicts the number fraction of galaxies where outflow is detected in bins of galaxy SFR. 
The black line in the latter five panels is the best-fit relation between $\dot{M}_{\rm out}$ and SFR. 
Details in \S\ref{sec-res-Flow-Mdot}. 
} 
\label{fig-Mdot-vs-SFR} 
\end{figure*} 
%%%%%%%%%%%%%%%%%%%%%%%%%%%%%%%%%%%%%%%%%%%%%%%%%%%%%%%%%%%%%%%%%%%%%%%% 

The mass outflow rate as a function of SFR of galaxies for the six remaining simulations 
is plotted in Fig.~\ref{fig-Mdot-vs-SFR}. 
The format is same as the bottom row of Fig.~\ref{fig-Mdot-vs-SFR-3runs-2cutoff}, where 
gas is selected above the escape velocity $|v_{z, i}| > v_{\rm esc}$. 
The plotting colour depicts the number fraction of galaxies where outflow is detected in bins of galaxy SFR. 

The no-wind run {\it E35nw} presents outflow in a few high-SFR galaxies only, 
and $\dot{M}_{\rm out}$ shows a scatter having no relation with the SFR. 
The two Effective models with kinetic SN feedback (Fig.~\ref{fig-Mdot-vs-SFR-3runs-2cutoff}) 
display a weak positive correlation of $\dot{M}_{\rm out}$ and galaxy SFR, 
having a value of slope $0.86$, which is flatter than the MUPPI models. 
{\it E25cw} has a larger scatter than {\it E35rvw}. 

The six MUPPI runs (five in Fig.~\ref{fig-Mdot-vs-SFR} and one in Fig.~\ref{fig-Mdot-vs-SFR-3runs-2cutoff}) 
exhibit a prominent positive correlation of $\dot{M}_{\rm out}$ with galaxy SFR, 
having slope values between $1$ and $1.2$. 
$\dot{M}_{\rm out}$ rises from $0.08 M_{\odot} /$ yr at low-SFR 
to $50 M_{\odot} /$ yr at high-SFR. 
All the runs present some scatter, 
with $\dot{M}_{\rm out}$ varying up to $10$ times for the same SFR values. 

% This positive correlation is stronger and relatively tighter than 
% that found between $v_{\rm out}$ and SFR in \S\ref{sec-res-Flow-Vel}. 
% , and {\it E35rvw} a tighter correlation than {\it E25cw}. 

\subsubsection{Mass Escape at Virial versus Galaxy Radius} 
\label{sec-res-Flow-Mdot-Rvir-vs-Rgal} 

%%%%%%%%%%%%%%%%%%%%%%%%%%%%%%%%%%%%%%%%%%%%%%%%%%%%%%%%%%%%%%%%%%%%%%%%%%%%%%%%%%%%%%%%%% 
%%%%%%%%%%%%%%%%%%%%%%%%%%%%%%%%%%%%%%%%%%%%%%%%%%%%%%%%%%%%%%%%%%%%%%%%%%%%%%%%%%%%%%%%%% 
% 
% TABLE 2 

\begin{table*} 
\begin{minipage}{10cm} 
\caption{ 
Number statistics of outflows in run {\it M50std}   % Muppi36 
at $z = 2$ using different techniques, described in \S\ref{sec-res-Flow-Mdot-Rvir-vs-Rgal}. 
Flows measured by tracking gas particles 
having a radial velocity higher than the escape: $|v_r| > v_{\rm esc}(r)$. 
Column 1: Method description. 
Column 2: $N_{\rm outflow}$ = Number of galaxies where outflow is detected. 
Column 3: Fraction $f_{\rm outflow} = N_{\rm outflow} / N_{\rm central}$. 
} 

\label{Table-Rgal-vs-Rvir} 
\begin{tabular}{ccc} 

\hline 
Method & $N_{\rm outflow}$ & $f_{\rm outflow}$ \\ 
\hline 

At $R_{\rm gal}$ using $|v_r| > v_{\rm esc}(R_{\rm gal})$, in a cylinder & $1842$ & $0.93$ \\ 
At $R_{\rm gal}$ using $|v_r| > v_{\rm esc}(R_{\rm gal})$, in a sphere   & $1936$ & $0.97$ \\ 
At $R_{\rm vir}$ using $|v_r| > v_{\rm esc}(R_{\rm vir})$, in a sphere   & $1734$ & $0.87$ \\ 

\hline 
\end{tabular} 

\end{minipage} 
\end{table*} 

%%%%%%%%%%%%%%%%%%%%%%%%%%%%%%%%%%%%%%%%%%%%%%%%%%%%%%%%%%%%%%%%%%%%%%%%%%%%%%%%%%%%%%%%%% 
%%%%%%%%%%%%%%%%%%%%%%%%%%%%%%%%%%%%%%%%%%%%%%%%%%%%%%%%%%%%%%%%%%%%%%%%%%%%%%%%%%%%%%%%%% 

We compare outflow measurements at $R_{\rm gal}$ using the cylinder and the sphere techniques, 
which reveal us their shapes. 
During propagation an outflow could be slowed down by hydrodynamical interactions with neighbouring gas. 
In order to estimate what mass fraction can escape the halo gravitational potential, 
we compare the rates by measuring outflow at the galaxy radius $R_{\rm gal}$ 
and that at the virial radius $R_{\rm vir}$. 

Here we present results of run {\it M50std} at $z = 2$ by analysing $N_{\rm central} = 1986$ 
central galaxies, among a total of $2688$ galaxies with $M_{\star} > 10^{9} M_{\odot}$. 
Table~\ref{Table-Rgal-vs-Rvir} lists the number of outflows detected 
using different methods at two radii: 
$r = R_{\rm gal}$ in the first two rows, and $r = R_{\rm vir}$ in the third row. 
We find that the fraction of galaxies where outflow is detected at $R_{\rm gal}$ 
is $f_{\rm outflow} = 0.93$ using the cylinder method, and rises to $0.97$ using the sphere technique. 
We furthermore find that the outflow detection fraction 
decreases from $f_{\rm outflow} = 0.97$ at $R_{\rm gal}$, to $0.87$ at $R_{\rm vir}$. 
The reduction factor is small; among the outflows which escape the galaxy, 
$\sim 90 \%$ can escape the halo as well. 

% \dot{M}_{\rm in}  = \sum_{i=1}^{n_{\rm in}}  \frac{m_i |v_{r, i}|} {h_{\rm sph}} 

%%%%%%%%%%%%%%%%%%%%%%%%%%%%%%%%%%%%%%%%%%%%%%%%%%%%%%%%%%%%%%%%%%%%%%%% 
% FIGURE 8 
\begin{figure*} 
\centering 
\includegraphics[width = 0.8 \linewidth]{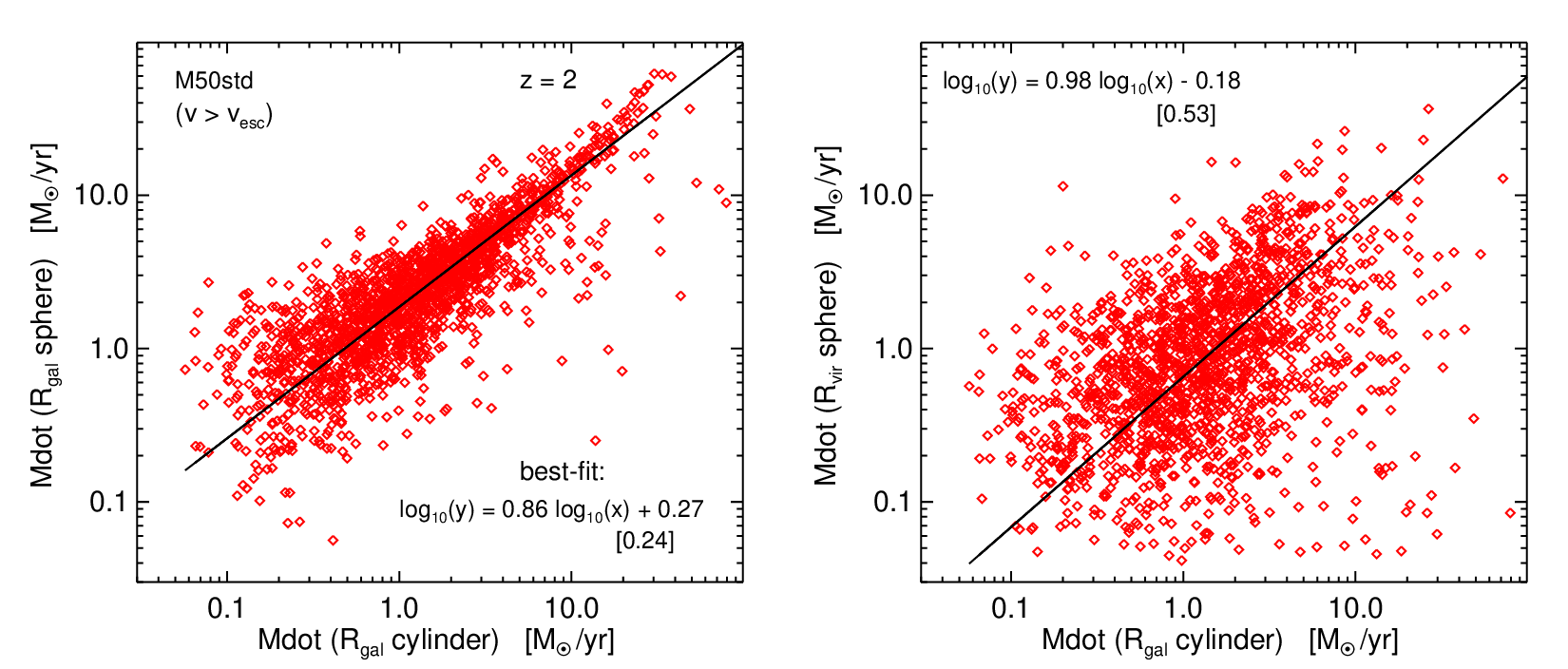} 
\caption{ 
Mass outflow rate in run {\it M50std} at $z = 2$ measured by different techniques; 
{\it left panel:} at the galaxy radius $R_{\rm gal}$ in a sphere (Eq.~\ref{eq-Mdot-Sph}) 
versus that at $R_{\rm gal}$ in a cylinder, 
and {\it right panel:} at the virial radius $R_{\rm vir}$ in a sphere 
versus that at $R_{\rm gal}$ in a cylinder. 
The black line shows the result of an outlier-resistant two-variable linear regression 
giving the best-fit relation between the plotted quantities. 
The fit coefficients and the ``standard deviation'' of the fit's residuals (within square brackets) 
are written inside the panels, using units same as the plotted axes. 
Described in \S\ref{sec-res-Flow-Mdot-Rvir-vs-Rgal}. 
} 
\label{fig-Mdot-At-Rvir-vs-Rgal} 
\end{figure*} 
%%%%%%%%%%%%%%%%%%%%%%%%%%%%%%%%%%%%%%%%%%%%%%%%%%%%%%%%%%%%%%%%%%%%%%%% 

Fig.~\ref{fig-Mdot-At-Rvir-vs-Rgal} presents the mass outflow rate in run {\it M50std} at $z = 2$ 
measured by different techniques; 
left panel: at the galaxy radius $R_{\rm gal}$ in a sphere versus that at $R_{\rm gal}$ in a cylinder, 
and right: at the virial radius $R_{\rm vir}$ in a sphere versus that at $R_{\rm gal}$ in a cylinder. 
Outflows are measured by selecting gas particles above 
the escape velocity $|v_{z, i}| > v_{\rm esc}(r)$ at the given radius. 
There is a substantial scatter in the right panel, with $\dot{M}_{\rm out}(R_{\rm vir})$ 
varying up to a few $10$'s at the same $\dot{M}_{\rm out}(R_{\rm gal})$; 
however a clear positive correlation is visible. 
The black line shows the result of an outlier-resistant two-variable linear regression. 
In the right panel, this gives the best-fit relation of the mass escape at the two radii: 
$\dot{M}_{\rm out}(R_{\rm vir}) = 0.66 \dot{M}_{\rm out}(R_{\rm gal})^{0.98}$. 
From the left panel and the detection number ratio in the previous paragraph 
we conclude that, the shape of the outflow is bi-polar in $\sim 95 \%$ galaxies. 

% as the red diamonds populate a scattered band parallel to the dotted line. 
% Our tests with both cylinder and sphere method indicate that the outflow is basically bipolar, 
% as expected, since the relevant sub-resolution models are constructed that way. 

\subsubsection{Mass Loading Factor} 
\label{sec-res-Flow-eta} 

%%%%%%%%%%%%%%%%%%%%%%%%%%%%%%%%%%%%%%%%%%%%%%%%%%%%%%%%%%%%%%%%%%%%%%%% 
% FIGURE 9 
\begin{figure*} 
\centering 
\includegraphics[width = 0.9 \linewidth]{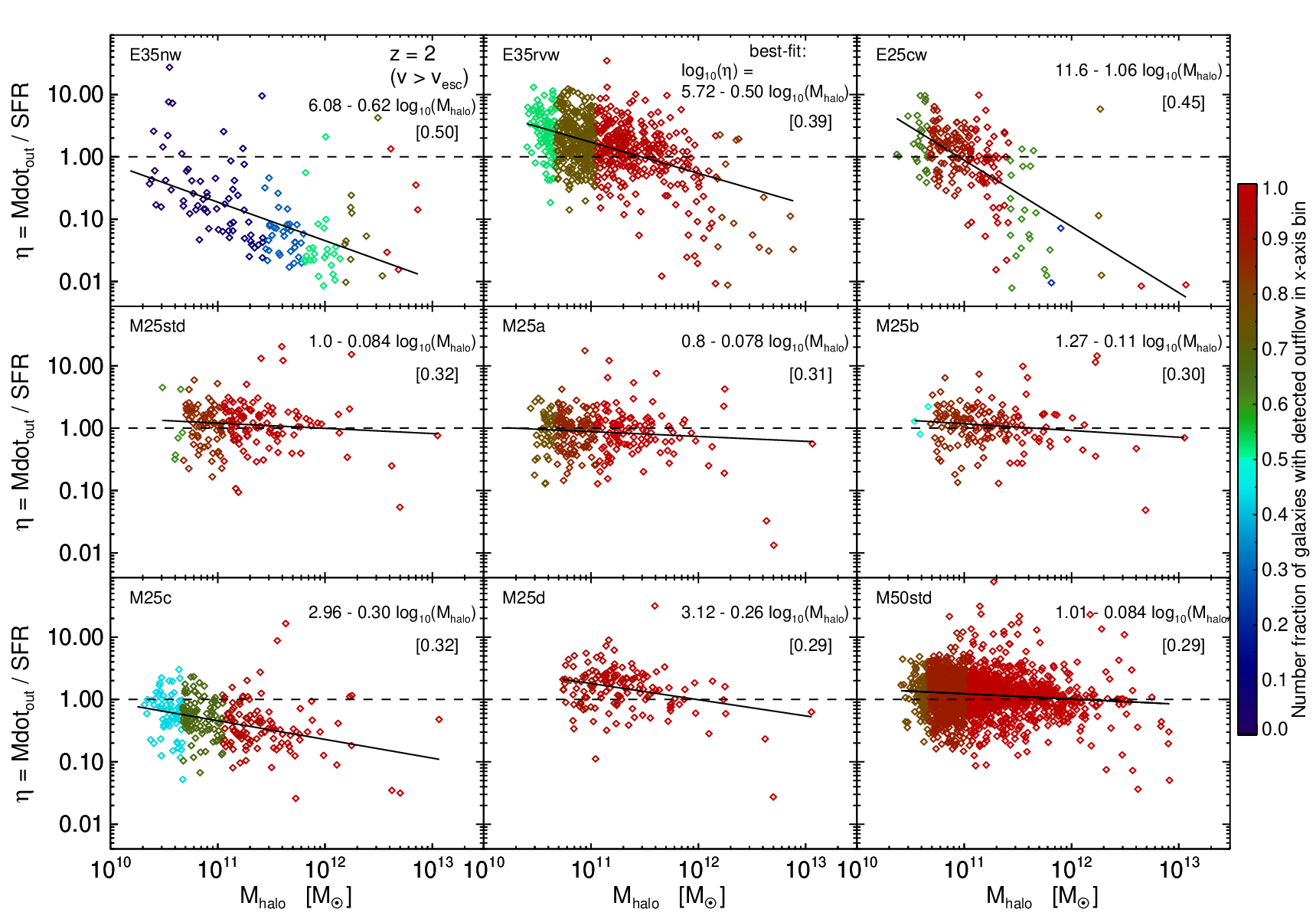} 
\caption{ 
Mass loading factor, $\eta$ from Eq.~(\ref{eq-FlowProp-eta}), as a function of halo mass of galaxies. 
Each of the nine panels shows one simulation. 
The format is the same as Fig.~\ref{fig-Mdot-vs-SFR}, where 
outflowing gas is selected above the escape velocity $|v_{z, i}| > v_{\rm esc}$. 
The plotting colour depicts the number fraction of galaxies where outflow is detected in bins of halo mass. 
Details in \S\ref{sec-res-Flow-eta}. 
} 
\label{fig-eta-vs-MFoF} 
\end{figure*} 
%%%%%%%%%%%%%%%%%%%%%%%%%%%%%%%%%%%%%%%%%%%%%%%%%%%%%%%%%%%%%%%%%%%%%%%% 

We compute the outflow mass loading factor $\eta$ by taking the ratio of 
the mass outflow rate (measured at $R_{\rm gal}$ using the bi-cylinder method) with galaxy SFR, 
\begin{equation} 
\label{eq-FlowProp-eta} 
\eta = \frac{ \dot{M}_{\rm out} } { {\rm SFR} }. 
\end{equation} 
Fig.~\ref{fig-eta-vs-MFoF} presents $\eta$ as a function of halo mass of galaxies 
for all nine simulations, using a format same as Fig.~\ref{fig-Mdot-vs-SFR}. 
The outflowing gas is selected above the escape velocity $|v_{z, i}| > v_{\rm esc}$. 
The plotting colour depicts the number fraction of galaxies where outflow is detected in bins of halo mass. 
All the runs exhibit a scatter, 
with $\eta$ varying up to factors of some-$10$'s at the same halo mass. 

The Effective model cases ({\it E35nw}, {\it E25cw} and {\it E35rvw} at higher masses) 
generate a negative correlation of $\eta$ with $M_{\rm halo}$. 
This is because two runs have kinetic SN feedback 
in the energy-driven outflow formalism (\S\ref{sec-num-EffMod}), 
where the wind kick velocity $v_w$ is fixed over all galaxy masses. 
The result turns out to be a strong wind for the low-mass galaxies, 
where a large amount of gas is expelled compared to SF, reaching $\eta \sim 10$. 
On the other hand, it is a weak wind for the high-mass galaxies, 
where little gas is expelled and $\eta \sim 0.01$. 
At lower masses ($M_{\rm halo} < 4 \times 10^{11} M_{\odot}$), 
run {\it E35rvw} shows a constant $\eta$ scattered between $0.4 - 10$ versus $M_{\rm halo}$. 

All the six MUPPI runs display an almost constant value of $\eta$ over the full range of $M_{\rm halo}$. 
The mass loading factors in most of the MUPPI galaxies lie between $\eta = 0.2 - 10$, 
with an average value of $\eta \sim 1$. 

% The no-wind run {\it E35nw} shows an outflow in the higher-mass galaxies only, 
% and it is weak with $\eta < 0.1$. 
% Only exception is the least efficient kinetic feedback case {\it M25c}, 
% where a weak negative correlation is seen and all the galaxies have $\eta \leq 1$. 

\subsubsection{Implications and Comparison with Theoretical Estimates} 
\label{sec-res-Flow-Compare} 

We find that the MUPPI model is able to produce galactic outflows whose 
{\it velocity and mass outflow rate correlates positively with 
global properties of the galaxy (halo mass, SFR)}. 
This is achieved with MUPPI {\it using fully local properties} of gas as input to the sub-resolution model. 
This trend, tracing some of the global properties of galaxy using local properties of gas, 
is found for the first time in cosmological simulations using sub-resolution models. 
We decipher that such trends arise from the details of the MUPPI sub-resolution model. 
Here star formation depends on local pressure through the molecular fraction of gas 
(Eq.~5 in \citealt{Murante10}), and the gas pressure is expected to depend on halo mass. 
The star forming gas undergoes kinetic feedback and produces galactic outflows. 
Therefore consequently the outflow properties become dependent on 
halo mass and other global properties, through the local gas pressure. 

On the other hand, the Effective model imparts a constant velocity $v_w$ to the gas, 
and a constant mass-loading $\eta = 2$, as input to the sub-resolution recipe. 
Consequently, the measured outflow speed is seen to be constant in the Effective model. 
However it produces a trend of $\eta$ decreasing (from a value of $10$ to $0.01$) with halo mass, 
and $\eta = 2$ is only achieved for a narrow range of galaxy masses. 

The results we obtain here by computing outflows in galaxies extracted from 
cosmological hydrodynamical simulations, can be compared to theoretical estimates 
of the MUPPI sub-resolution model. 
At variance with other kinetic wind recipes in the literature, 
in MUPPI neither the wind mass loading factor nor the wind velocity are given as input quantities. 
Nevertheless, typical values of these can be estimated 
using certain hypothesis about the average properties  of the gas, 
as done by \citet{Murante14} in their Section 3.3. 
With the default parameter choices, the theoretical estimates are 
(Eq.~\ref{eq-Muppi-eta} and Eq.~\ref{eq-Muppi-Vwind2} in \S\ref{sec-num-Muppi}): 
$\eta_{\rm th} \simeq 1.5$, and 
the mass-weighted average wind velocity $v_{\rm wind, th} \simeq 600$ km/s. 
However the exact values depend on the local dynamical time, hence in turn on the local gas properties. 

The Muppi runs {\it M25std} and {\it M50std} yield outflow mass loading factor between 
$\eta = 0.3 - 8$, with an average value of $\eta \simeq 1.2$, close to $\eta_{\rm th} \simeq 1.5$. 
The outflow velocity shows a positive correlation with galaxy mass and SFR, 
$v_{\rm out}$ rising from $300$ km/s to $600$ km/s. 
Hence the simulation $v_{\rm out}$ agrees with the theoretical estimate in higher-mass galaxies, 
but at lower-masses the outflow speed is smaller than $v_{\rm wind, th} \simeq 600$ km/s. 

%%%%%%%%%%%%%%%%%%%%%%%%%%%%%%%%%%%%%%%%%%%%%%%%%%%%%%%%%%%%%%%%%%%%%%%% 
% FIGURE 10 
\begin{figure*} 
\centering 
\includegraphics[width = 0.5 \linewidth]{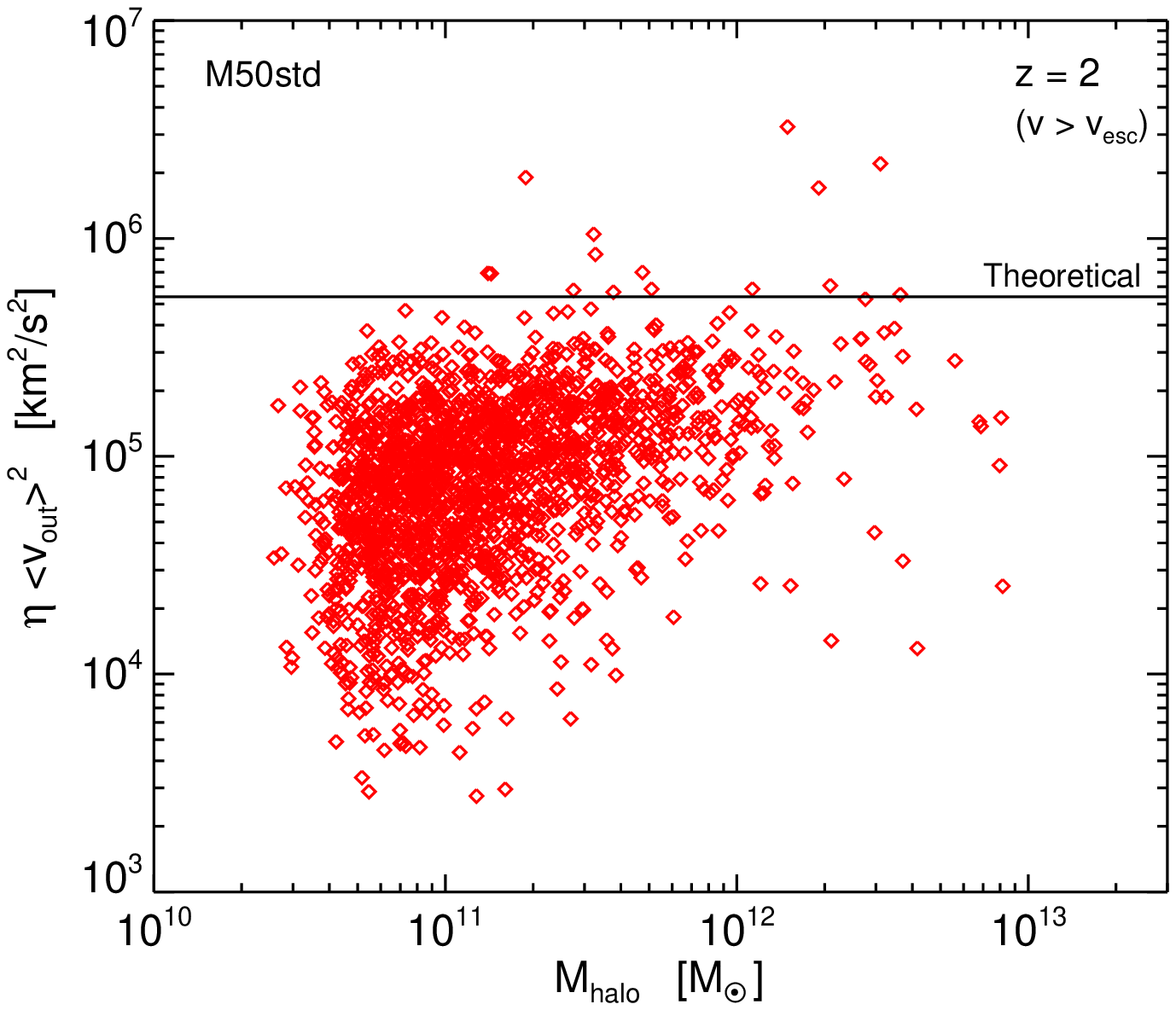} 
\caption{ 
Product of mass loading factor and outflow velocity squared, $\eta ~ \langle v_{\rm out} \rangle^2$, 
as a function of halo mass of galaxies, in run {\it M50std}. 
Outflows measured by selecting gas particles above the escape velocity $|v_{z, i}| > v_{\rm esc}$. 
The horizontal {\it black solid} line marks the the theoretical estimate 
$\eta_{\rm th} v_{\rm wind, th}^2$ \citep{Murante14}. Description in \S\ref{sec-res-Flow-Compare}. 
} 
\label{fig-eta-times-Vel2-vs-MFoF} 
\end{figure*} 
%%%%%%%%%%%%%%%%%%%%%%%%%%%%%%%%%%%%%%%%%%%%%%%%%%%%%%%%%%%%%%%%%%%%%%%% 

The product of mass loading factor and outflow velocity squared is predicted to be 
a constant in the MUPPI model \citep{Murante14}: 
$\eta_{\rm th} v_{\rm wind, th}^2 = f_{\rm fb,kin} E_{\rm SN} / M_{\rm\star,SN}$. 
This product for the simulated galaxies $\eta ~ \langle v_{\rm out} \rangle^2$ 
as a function of halo mass is plotted in Fig.~\ref{fig-eta-times-Vel2-vs-MFoF} for run {\it M50std}. 
Here outflows are measured by selecting gas particles above the escape velocity $|v_{z, i}| > v_{\rm esc}$. 
The horizontal {\it black solid} line marks the theoretical estimate 
$\eta_{\rm th} v_{\rm wind, th}^2 = 5.4 \times 10^5$ km$^2$/s$^2$. 
The simulation results are lower than the theoretical prediction, especially in the less-massive galaxies. 
This is because the prediction assumes certain average properties of the gas. 
We in turn conclude that only a fraction $1/3$rd of the deposited SN energy is used to drive an outflow; 
the rest being dispersed and radiated away through hydrodynamical interactions 
and by giving energy to slow particles.

\subsection{Redshift Evolution of Outflows: from $z = 5$ to $z = 1$} 
\label{sec-res-zEvol} 

%%%%%%%%%%%%%%%%%%%%%%%%%%%%%%%%%%%%%%%%%%%%%%%%%%%%%%%%%%%%%%%%%%%%%%%% 
% FIGURE 11 
\begin{figure*} 
\centering 
\includegraphics[width = 0.8 \linewidth]{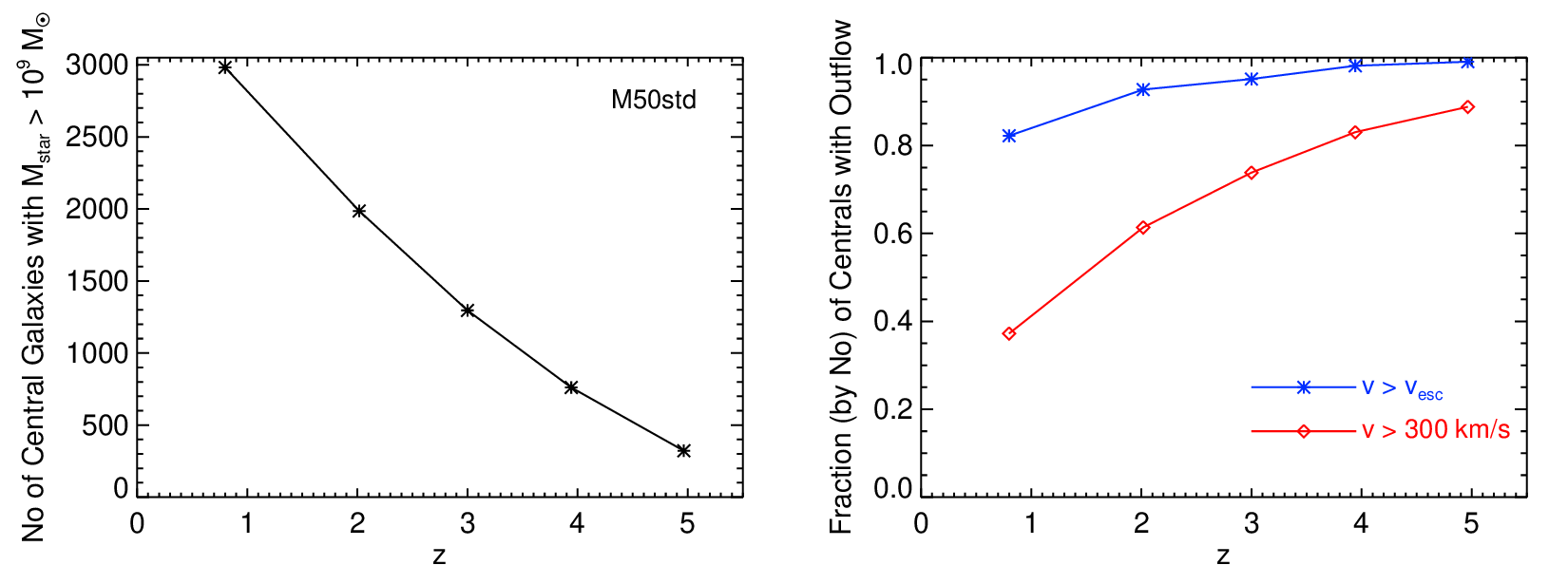} 
\caption{ 
Number of central galaxies in run {\it M50std} having a stellar mass 
$M_{\star} > 10^{9} M_{\odot}$ in the {\it left panel}, 
and number fraction of these centrals where outflow is detected at the {\it right}, 
as a function of redshift. 
The two curves in the right panel indicate different methods of 
measuring outflow (\S\ref{sec-num-FlowMeasureTech}), by selecting gas above: 
a constant limiting speed $|v_{z, i}| > 300$ km/s ({\it diamonds, red} curve), 
and the escape velocity $|v_{z, i}| > v_{\rm esc}$ ({\it asterisks, blue} curve). 
Description in \S\ref{sec-res-zEvol}. 
} 
\label{fig-FracOutflow-vs-z} 
\end{figure*} 
%%%%%%%%%%%%%%%%%%%%%%%%%%%%%%%%%%%%%%%%%%%%%%%%%%%%%%%%%%%%%%%%%%%%%%%% 

The outflow detection fraction of run {\it M50std} as a function of redshift 
is plotted in Fig.~\ref{fig-FracOutflow-vs-z}. 
Left panel shows the absolute number of central galaxies in simulation volume 
having a stellar mass $M_{\star} > 10^{9} M_{\odot}$. 
Right panel illustrates the number fraction of these centrals where outflow is detected 
(defined in Table~\ref{Table-Rgal-vs-Rvir}), 
using the cylindrical volume methodology of \S\ref{sec-num-FlowMeasureTech}. 
When outflow is measured by selecting gas above the escape velocity 
($|v_{z, i}| > v_{\rm esc}$, {\it asterisks, blue} curve), 
the outflow detection fractions are high at all epochs, 
reducing gradually from $f_{\rm outflow} = 0.99$ at $z = 5$, to $0.8$ at $z \sim 1$. 
When measured above a constant limiting speed ($|v_{z, i}| > 300$ km/s, {\it diamonds, red} curve), 
$f_{\rm outflow}$ decreases from $0.8$ at $z = 4$, to $0.4$ at $z \sim 1$. 

% Inflow detection fractions ({\it back plus symbols}) are always small, 
% $f_{\rm inflow} < 0.17$, but present a similar redshift trend as the outflows. 
% $0.6$ at $z = 2$, and 

%%%%%%%%%%%%%%%%%%%%%%%%%%%%%%%%%%%%%%%%%%%%%%%%%%%%%%%%%%%%%%%%%%%%%%%% 
% FIGURE 12 
\begin{figure*} 
\centering 
\includegraphics[width = 0.9 \linewidth]{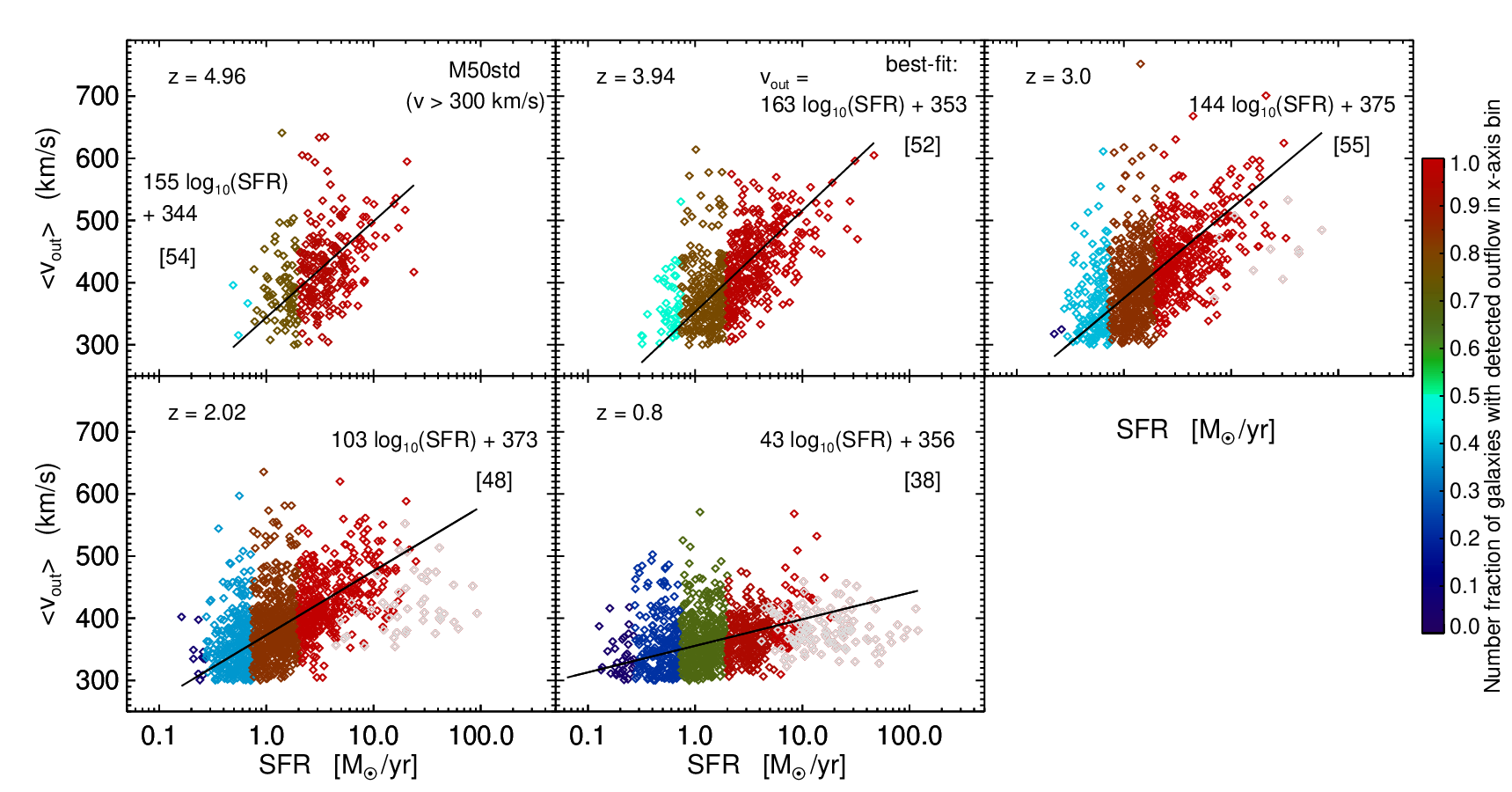} 
\caption{ 
Redshift evolution of outflow velocity mass-weighted average, 
$\langle v_{\rm out} \rangle$ from Eq.~(\ref{eq-FlowProp-MassAvg}), as a function of SFR of galaxies. 
Here outflowing gas is selected above a constant limiting speed $|v_{z, i}| > 300$ km/s. 
Each point is one galaxy of simulation run {\it M50std}, 
and the five panels show epochs of $z \sim 5, 4, 3, 2, 0.8$. 
The plotting colour depicts the number fraction of galaxies where outflow is detected in bins of galaxy SFR. 
The grey points mark galaxies more massive than $M_{\rm halo, lim} = 1.3 \times 10^{12} M_{\odot}$, 
where the outflows might not escape the halo potential (\S\ref{sec-num-FlowMeasureTech}). 
Discussed in \S\ref{sec-res-zEvol}. 
} 
\label{fig-zEvol-Vout-vs-SFR} 
\end{figure*} 
%%%%%%%%%%%%%%%%%%%%%%%%%%%%%%%%%%%%%%%%%%%%%%%%%%%%%%%%%%%%%%%%%%%%%%%% 

The redshift evolution of outflow velocity in simulation run {\it M50std} 
as a function of SFR of galaxies is plotted in Fig.~\ref{fig-zEvol-Vout-vs-SFR}. 
It shows the mass-weighted average, 
$\langle v_{\rm out} \rangle$ from Eq.~(\ref{eq-FlowProp-MassAvg}) in \S\ref{sec-num-FlowMeasureTech}. 
Each point is one galaxy, and the five panels show different redshifts. 
The remaining plotting format is the same as in Fig.~\ref{fig-Vout-vs-SFR}. 
The correlation between $v_{\rm out}$ and SFR is positive at all the explored epochs: 
steeper at earlier times, and becomes flatter at later epochs. 
The best-fit slope of $v_{\rm out}$ (km/s) versus log$_{10}$ (SFR/$M_{\odot}$ yr$^{-1})$ 
are: $155, ~ 163, ~ 144, ~ 103, ~ 43$ at $z \sim 5, ~ 4, ~ 3, ~ 2, ~ 0.8$.

%%%%%%%%%%%%%%%%%%%%%%%%%%%%%%%%%%%%%%%%%%%%%%%%%%%%%%%%%%%%%%%%%%%%%%%% 
% FIGURE 13 
\begin{figure*} 
\centering 
\includegraphics[width = 0.9 \linewidth]{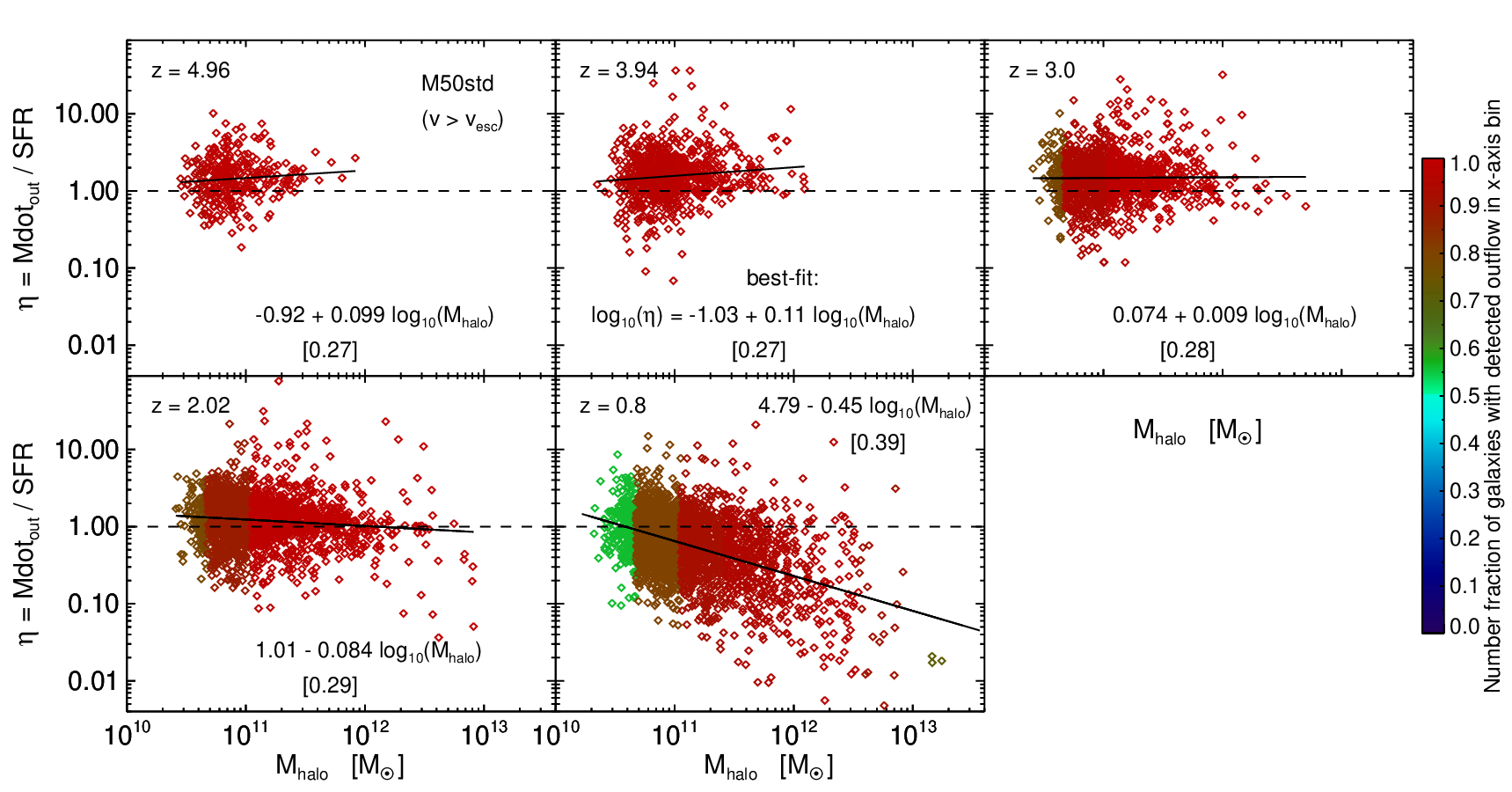} 
\caption{ 
Redshift evolution of mass loading factor, $\eta$ from Eq.~(\ref{eq-FlowProp-eta}), 
as a function of halo mass of galaxies. 
The format is the same as Fig.~\ref{fig-zEvol-Vout-vs-SFR}, 
showing simulation {\it M50std} at different epochs. 
Except here the outflowing gas is selected above the escape velocity $|v_{z, i}| > v_{\rm esc}$. 
The plotting colour depicts the number fraction of galaxies where outflow is detected in bins of halo mass. 
Details in \S\ref{sec-res-zEvol}. 
} 
\label{fig-zEvol-eta-vs-MFoF} 
\end{figure*} 
%%%%%%%%%%%%%%%%%%%%%%%%%%%%%%%%%%%%%%%%%%%%%%%%%%%%%%%%%%%%%%%%%%%%%%%% 

The redshift evolution of mass loading factor of galaxies in run {\it M50std} 
as a function of halo mass is plotted in Fig.~\ref{fig-zEvol-eta-vs-MFoF}. 
It shows $\eta$ from Eq.~(\ref{eq-FlowProp-eta}), % using a format same 
as in Fig.~\ref{fig-zEvol-Vout-vs-SFR}, presenting the simulation at different epochs. 
In this figure the plotting colour depicts the 
number fraction of galaxies where outflow is detected in bins of halo mass. 
The value of $\eta$ exhibits a scatter at all the epochs, 
varying up to factors of $10$ at the same halo mass. 
Earlier on between $z = 2 - 5$, 
the galaxies have an almost constant value of $\eta$ over the full range of $M_{\rm halo}$. 
The mass loading factors lie within $\eta = 0.1 - 20$, 
with an average value decreasing with the passage of time. 
At later epochs $z = 0.8$, 
$\eta$ displays a negative correlation with $M_{\rm halo}$ at the high-mass end.  

The most important factor causing these redshift evolution of the outflow properties 
is the decrease of SFR at low-$z$ (\S\ref{sec-res-SFRD}). 
As seen in Fig.~\ref{fig-SFRD}, the global SFRD in run {\it M50std} reaches a maximum between 
$z = 2 - 4$ in the form of a plateau, with a steeper reduction of SFRD at earlier and later redshifts. 
This high SFR activity drives strong outflows in a greater fraction of galaxies, 
and causes the positive correlation of $v_{\rm out}$ with SFR, as well as the constant-$\eta$ trend. 
Later at $z \leq 1$, SFR in galaxies reduce due to overall gas depletion, 
hence the driven outflows become weaker and rarer, and the outflow correlations are lost. 

% and all the galaxies have $\eta \leq 1$. 

% The interplay of multiple factors cause these redshift evolution of the outflow properties. 
% There is also a change of galaxy morphology with time, which contributes to the redshift evolution. 
% More galaxies turn elliptical / bulge-dominated at late times, causing a reduction of outflow in them. 

\subsection{Comparison of Outflows with Observations and Other Models} 
\label{sec-res-Compare} 

% Their Figure 7 (bottom panel) shows that the outflow terminal velocities are always 
% $2 - 3$ times larger than the galactic rotation speed. 

Observational signatures of galactic outflows mostly comprise of single-galaxy detections. 
As a recent example at high redshift, 
\citet{Crighton14} observed absorbing gas clumps in a $z = 2.5$ galaxy CGM, 
produced by a wind with a mass outflow rate of $\sim 5 M_{\odot} /$yr. 
% of high metallicity $(0.1 - 0.6) Z_{\odot}$, cool temperature ($10^{4}$ K), 
% and densities $10^{-3} - 10^{-2}$ cm$^{-3}$; 
% however the clumps are of extremely small sizes ($< 100 - 500$ pc). 
% Presenting a multi-wavelength integral field spectroscopic study, 
At low redshift, \citet{Cazzoli14} detected kpc-scale neutral gas 
outflowing perpendicular to the disk of a $z = 0.01$ % luminous infrared 
galaxy, at a rate $\sim 48 M_{\odot} /$ yr, with a global mass loading factor $\eta \approx 1.4$. 
% however the 2D distribution of the ongoing SF as traced by the H$\alpha$ emission map 
% suggests a much larger value of $\eta$ associated with the inner $r < 200$ pc regions, 
% where the current observed SF represents only 3 percent of the total. 
Such values of mass outflow rate (a few to 10's $M_{\odot} /$yr) and $\eta$ 
lie within our simulation result range (\S\ref{sec-res-Flow-Mdot}, \S\ref{sec-res-Flow-eta}). 

Only over the last few years, observations of outflows in galaxy populations have been possible, 
and subsequent derivation of systematic trends. 
\citet{Grimes09} detected starburst-driven galactic winds of temperature $10^{5.5}$ K 
in the absorption spectra of 16 local galaxies, 
% covering 3 orders of magnitude in SFR and 2 orders of magnitude in stellar mass 
and found that $v_{\rm out}$ increases with both the SFR and the SFR per unit stellar mass. 
% agrees with a galactic wind, driven by the population of massive stars. 
In a spectroscopic catalogue of 40 luminous starburst galaxies at $0.7 \leq z \leq 1.7$, 
\citet{Banerji11} inferred the presence of large-scale outflowing gas, 
with $v_{\rm out} \propto {\rm SFR}^{0.3}$. 
% which is the same as the local envelope seen in lower redshift ultraluminous infrared galaxies. 
Analysing the cool outflow around galaxies at $1 \leq z \leq 1.5$, % via Mg II absorption, 
\citet{Bordoloi13} found that $v_{\rm out}$ (ranging between $200 - 300$ km/s) increases steadily 
with increasing SFR and stellar mass, and the wind is bipolar in geometry for disk galaxies. 
% The two highest velocity correspond to AGNs. 
% Conclude that massive galaxies are characterised by significantly higher velocity flows than 
% the typical Lyman-break galaxies at $z \sim 3$. 
At low redshifts, \citet{Martin05} observed a positive correlation of 
outflow speed with galaxy mass in ultraluminous infrared galaxies at $z = 0.042 - 0.16$. 
Our simulations predict a positive correlation of $v_{\rm out}$ with galaxy SFR 
and halo mass (\S\ref{sec-res-Flow-Vel}), especially at $z \geq 2$ (\S\ref{sec-res-zEvol}); 
hence consistent with all these observations. 

% The Muppi model trend of positive correlation of $v_{\rm out}$ with galaxy SFR 

Some galaxies in our simulations have a large $v_{\rm out} \sim 500 - 700$ km/s. 
This is in agreement with the high-velocity ($480 - 651$ km/s) outflow 
observations by \citet{Karman14} at $z \sim 3$ in the UV spectra of massive galaxies. 
Our simulated trend that the outflow detection fraction decreases from $z = 3$ to $z = 0.8$ 
(\S\ref{sec-res-zEvol}), 
is consistent with that observed by \citet{Karman14}: 
the incidence of high-$v_{\rm out}$ outflows ($40\%$) is much higher at $z \sim 3$ massive galaxies 
than those at $z < 1$; which is justified by the 
powerful SF and nuclear activity that most massive galaxies display at $z > 2$. 

An earlier simulation work by \citet{Oppenheimer08} yielded galactic outflows with 
faster wind speeds at high-$z$ and slower winds at low-$z$. 
This is in accord with our simulated redshift evolution of $v_{\rm out}$, 
which becomes weaker at $z \leq 2$    % and the positive correlation with SFR is lost 
(Fig.~\ref{fig-zEvol-Vout-vs-SFR}, \S\ref{sec-res-zEvol}). 
This trend, and the decrease of outflow detection fraction in our simulations from $z = 3$ to later epochs, 
are in agreement with observations indicating that 
high-$z$ ($z \sim 1 - 3$) galaxies almost ubiquitously reveal signatures of powerful winds 
\citep{Veilleux05}, than those at lower-$z$. 
We could  not compare with the recent $Illustris$ and $EAGLE$ simulations, 
because such outflow analysis has not been presented there. 
Our results are consistent with the recent work by \citet{Yabe14}, 
who employed a simple analytic model on an observational spectroscopic sample, 
and found that the gas outflow rate of star-forming galaxies 
decreases with decreasing redshift from $z \sim 2.2$ to $z \sim 0$. 
% which implies the higher activity of gas flow process at higher redshift. 

\subsection{Radial Profiles of Gas Properties in Galaxies and their CGM at $z = 2$} 
\label{sec-res-Radial-Profile} 

The radial profiles of gas properties around galaxy centers at $z = 2$ are presented in this section, 
considering all the galaxies (both centrals and satellites obtained by {\it SubFind}). 
Each property is shown as a function of galaxy radius, 
or distance from the location of subhalo potential minimum. 
All the non-wind gas particles lying inside a distance $R_{\rm lim}$ from the center are counted. 
The four panels denote total subhalo mass ($M_{\rm subhalo} / M_{\odot}$) ranges: 
$4 \times 10^{9} - 10^{10}$ (top-left) with number of subhalos in the different simulations 
within the range $N_{\rm subhalo} = 1977 - 17634$; 
$10^{10} - 10^{11}$ (top-right) having $1200 - 9801$ subhalos; 
$10^{11} - 10^{12}$ (bottom-left) with $85 - 771$ subhalos; and 
$10^{12} - 10^{13}$ (bottom-right) with $6 - 42$ subhalos. 

All the subhalos within each mass range are stacked, and the plotted solid curves denote 
the median value in radial bins for each run. 
The shaded areas mark the region between $25$th and $75$th percentiles 
in runs {\it E25cw} (black curve) and {\it M25std} (red curve). 
It shows the typical scatter at a given radius, 
since galaxies do not have spherically-symmetric properties in general. 
The vertical dashed line is the virial radius $R_{200}$ % in comoving coordinates 
for the following masses in the panels from top-left: 
$M_{\rm subhalo} / M_{\odot} = 6 \times 10^{9}$, $3 \times 10^{10}$, 
$3 \times 10^{11}$, and $3 \times 10^{12}$, 
where the exact values are $R_{200} = 59, 101, 218$, and $470$ kpc respectively. 
The outer plotting radius $R_{\rm lim}$ is chosen to be twice the virial radius, 
or $R_{\rm lim} = 2 ~ R_{200}$. 

Wind particles, or gas particles which have recently received 
a velocity kick from kinetic SN feedback and are being decoupled from hydrodynamic interactions 
(\S\ref{sec-num-EffMod}, \S\ref{sec-num-Muppi}) 
are excluded while computing the profiles. 

% The wind-phase gas typical of the effective SF model was seen as the cold ($< 10^4$ K), 
% dense ($> n_{\rm SF}$) tail in the bottom-right portion of the $[T - \rho]$ phase diagram 
% of run {\it E25cw} (Fig.~\ref{fig-Rho-vs-T} middle column, \S\ref{sec-res-T-rho}). 

% (analytical expression from Eq. \ref{eq-Mhalo}) % (DM + gas + star) 

\subsubsection{Density} 
\label{sec-res-Radial-Density} 

%%%%%%%%%%%%%%%%%%%%%%%%%%%%%%%%%%%%%%%%%%%%%%%%%%%%%%%%%%%%%%%%%%%%%%%% 
% FIGURE 14 
\begin{figure*} 
\centering 
\includegraphics[width = 1.0 \linewidth]{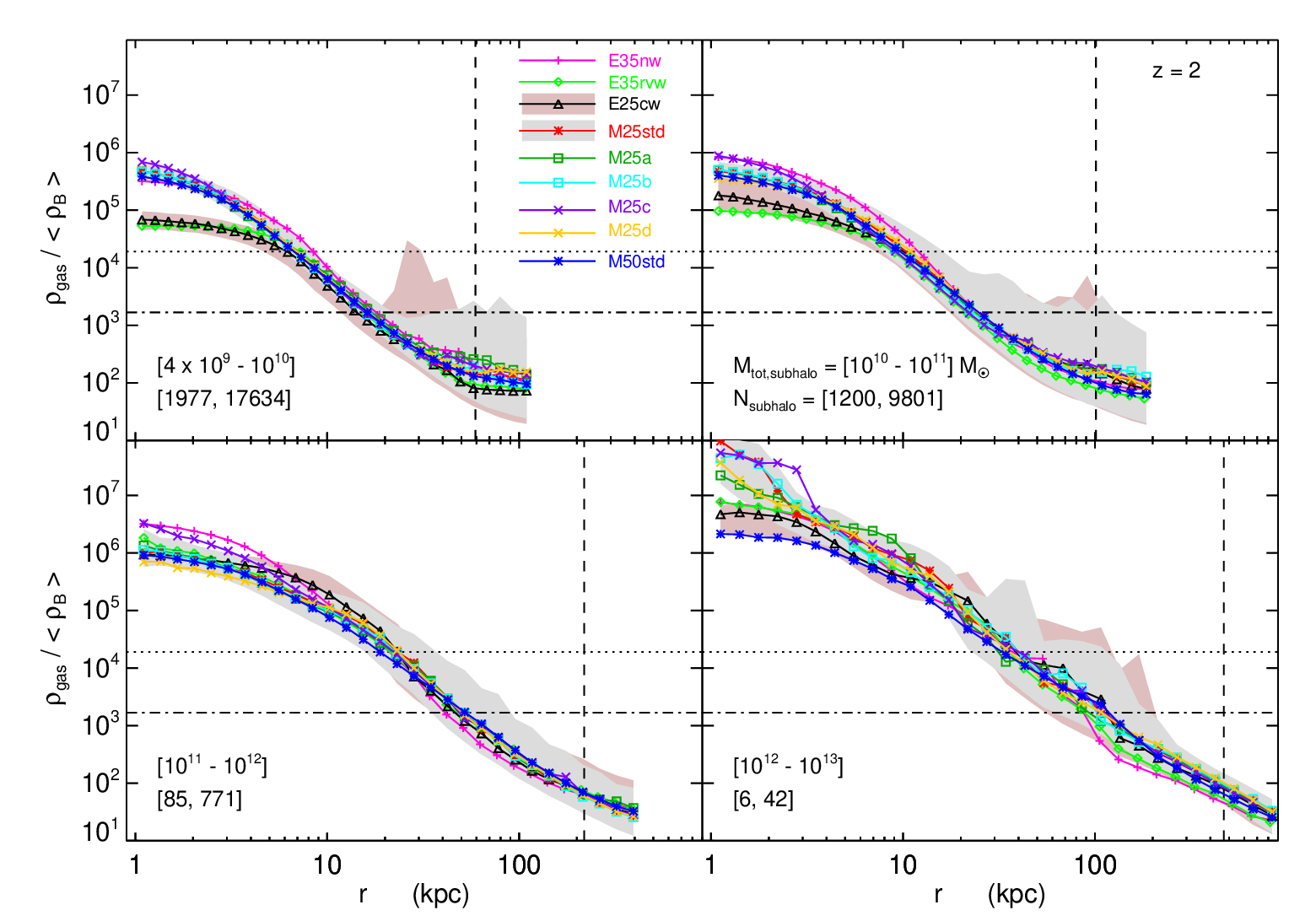} 
\caption{ 
Radial profiles of gas overdensity around galaxy centers at $z = 2$, 
with the different runs labelled by the colour and plotting symbol. 
Overdensity is plotted as a function of galaxy radius, 
for four total subhalo mass ranges in the four panels: 
$4 \times 10^{9} - 10^{10}$ ({\it top-left}), $10^{10} - 10^{11}$ ({\it top-right}), 
$10^{11} - 10^{12}$ ({\it bottom-left}), and $10^{12} - 10^{13}$ ({\it bottom-right}). 
All the subhalos within each mass range for each run (the number mentioned in the panels) 
are stacked over, and the plotted curve denotes the median value in a radial bin. 
The shaded areas enclose the $50$th percentiles 
above and below the median in runs {\it E25cw} (black curve) and {\it M25std} (red curve), 
showing the typical radial scatter. 
The {\it vertical dashed} line indicates the virial radius $R_{200}$ % in comoving coordinates, 
for $M_{\rm subhalo} / M_{\odot} = 6 \times 10^{9}$, $3 \times 10^{10}$, 
$3 \times 10^{11}$, and $3 \times 10^{12}$ in the panels from top-left respectively. 
The horizontal lines mark the SF threshold densities: 
$n_{\rm SF} = 0.13$ cm$^{-3}$ for the effective model as {\it dotted}, and 
$n_{\rm SF} = 0.01$ cm$^{-3}$ for the MUPPI model as {\it dot-dashed}. 
} 
\label{fig-rho-vs-R-MassBins} 
\end{figure*} 
%%%%%%%%%%%%%%%%%%%%%%%%%%%%%%%%%%%%%%%%%%%%%%%%%%%%%%%%%%%%%%%%%%%%%%%% 

The gas overdensity (ratio of density to the mean baryon density) 
radial profiles are plotted in Fig.~\ref{fig-rho-vs-R-MassBins}. 
Here the horizontal lines mark the SF threshold densities: 
$n_{\rm SF} = 0.13$ cm$^{-3}$ for the effective model (\S\ref{sec-num-EffMod}) as dotted, and 
$n_{\rm SF} = 0.01$ cm$^{-3}$ for the MUPPI model (\S\ref{sec-num-Muppi}) as dot-dashed. 
The gas denser than these thresholds in the respective models is forming stars. 
Within the approximate virial radius $r < R_{200}$, all the gas density profiles can be 
roughly described by two power laws with a break 
at an intermediate radius of $\sim 10$ kpc, dependent on halo mass and wind model. 

The inner parts $r < 10$ kpc of the two lower subhalo mass ranges 
($4 \times 10^{9} - 10^{10}$ and $10^{10} - 10^{11}$, top two panels) present notable differences: 
{\it E35rvw} and {\it E25cw} runs produce a lower density, by $10 - 30$ times, than the others. 
The model input wind speed is independent of halo mass: 
constant $350$ km/s in {\it E25cw}, and 
that dependent on galactocentric radius, $v_w(r)$, in {\it E35rvw}. 
Such a velocity is high enough to eject the gas away from the halo potential in low-mass galaxies, 
making the inner density smaller in these two runs. 
The central gas is not able to escape in cases {\it E35nw} (no-wind) which has no kinetic SN feedback, 
and {\it M25c} which has less efficient feedback, 
therefore producing the highest gas density at galaxy cores than the other runs. 

% (ratio of comoving density to the comoving mean baryon density, \S\ref{sec-res-T-rho}) 
% Whereas in high-mass galaxies, the velocity is not high enough to overcome the halo potential, 
% therefore the gas is not able to escape and retains back forming a high-density core. 
% two higher halo mass ranges ($10^{11} - 10^{12}$ and $10^{12} - 10^{13}$, right two panels), 

\subsubsection{Temperature} 
\label{sec-res-Radial-T} 

%%%%%%%%%%%%%%%%%%%%%%%%%%%%%%%%%%%%%%%%%%%%%%%%%%%%%%%%%%%%%%%%%%%%%%%% 
% FIGURE 15 
\begin{figure*} 
\centering 
\includegraphics[width = 1.0 \linewidth]{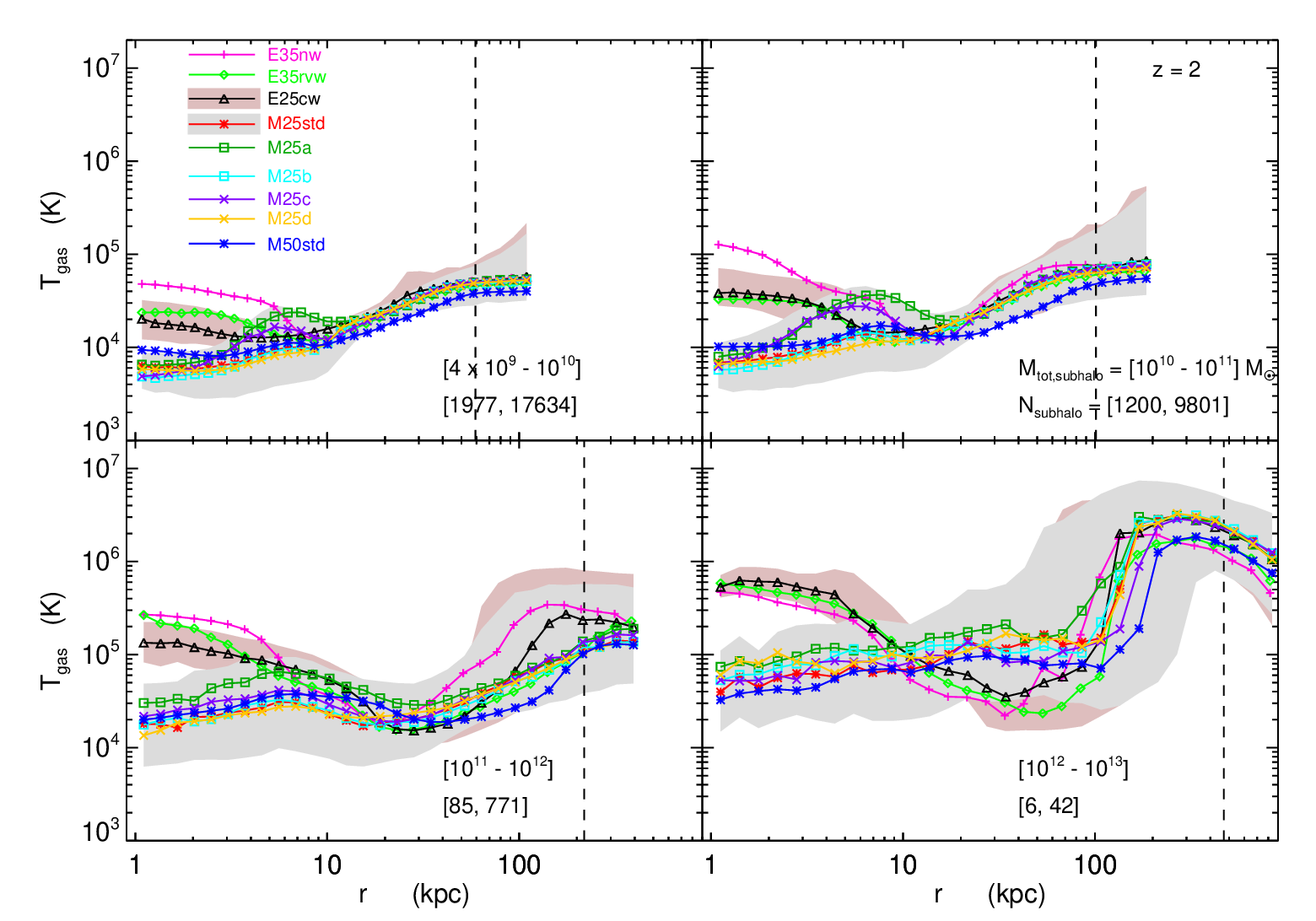} 
\caption{ 
Radial profiles of gas temperature around galaxy centers at $z = 2$, 
in a format same as Fig.~\ref{fig-rho-vs-R-MassBins}. 
} 
\label{fig-T-vs-R-MassBins} 
\end{figure*} 
%%%%%%%%%%%%%%%%%%%%%%%%%%%%%%%%%%%%%%%%%%%%%%%%%%%%%%%%%%%%%%%%%%%%%%%% 

The gas mass-weighted temperature radial profiles are presented in Fig.~\ref{fig-T-vs-R-MassBins}, 
in the same format as for density profiles. 
The average temperature has been used for those gas particles which are multiphase (star-forming). 
In all the effective model runs ({\it E35nw}, {\it E35rvw}, {\it E25cw}), 
the $T$-profiles in the inner parts $r \leq 6 - 30$ kpc of the galaxies 
follow the negative-sloped density-profiles (Fig.~\ref{fig-rho-vs-R-MassBins}). 
This region contains dense gas forming stars at galaxy centers. 
The central gas undergoing SF has a warm to high temperature 
($\sim 2 \times 10^{4} - 10^{7}$ K) by construction, 
as a result of following the SF effective equation of state.   % (\S\ref{sec-res-T-rho}). 
The MUPPI model produces a colder ($T \sim 5 \times 10^{3} - 10^{5}$ K) galaxy core 
at $r \leq 4 - 10$ kpc than the effective model. 

There is a change in $T$ slope in the outer parts, 
at $r \geq 15$ kpc in the top two panels, and 
at $r \geq (80 - 120)$ kpc in the bottom two panels; 
where the gas $T$ increases with radius, because of shock heating at galaxy outskirts. 
The $T$-profiles in the two higher subhalo mass ranges 
($10^{11} - 10^{12}$ and $10^{12} - 10^{13}$, bottom two panels) show a local peak at $200 - 300$ kpc, 
where the infalling gas collides with that in the halo and heats, 
to cool before reaching the center.

\subsubsection{Metallicity} 
\label{sec-res-Radial-Ztot} 

%%%%%%%%%%%%%%%%%%%%%%%%%%%%%%%%%%%%%%%%%%%%%%%%%%%%%%%%%%%%%%%%%%%%%%%% 
% FIGURE 16 
\begin{figure*} 
\centering 
\includegraphics[width = 1.0 \linewidth]{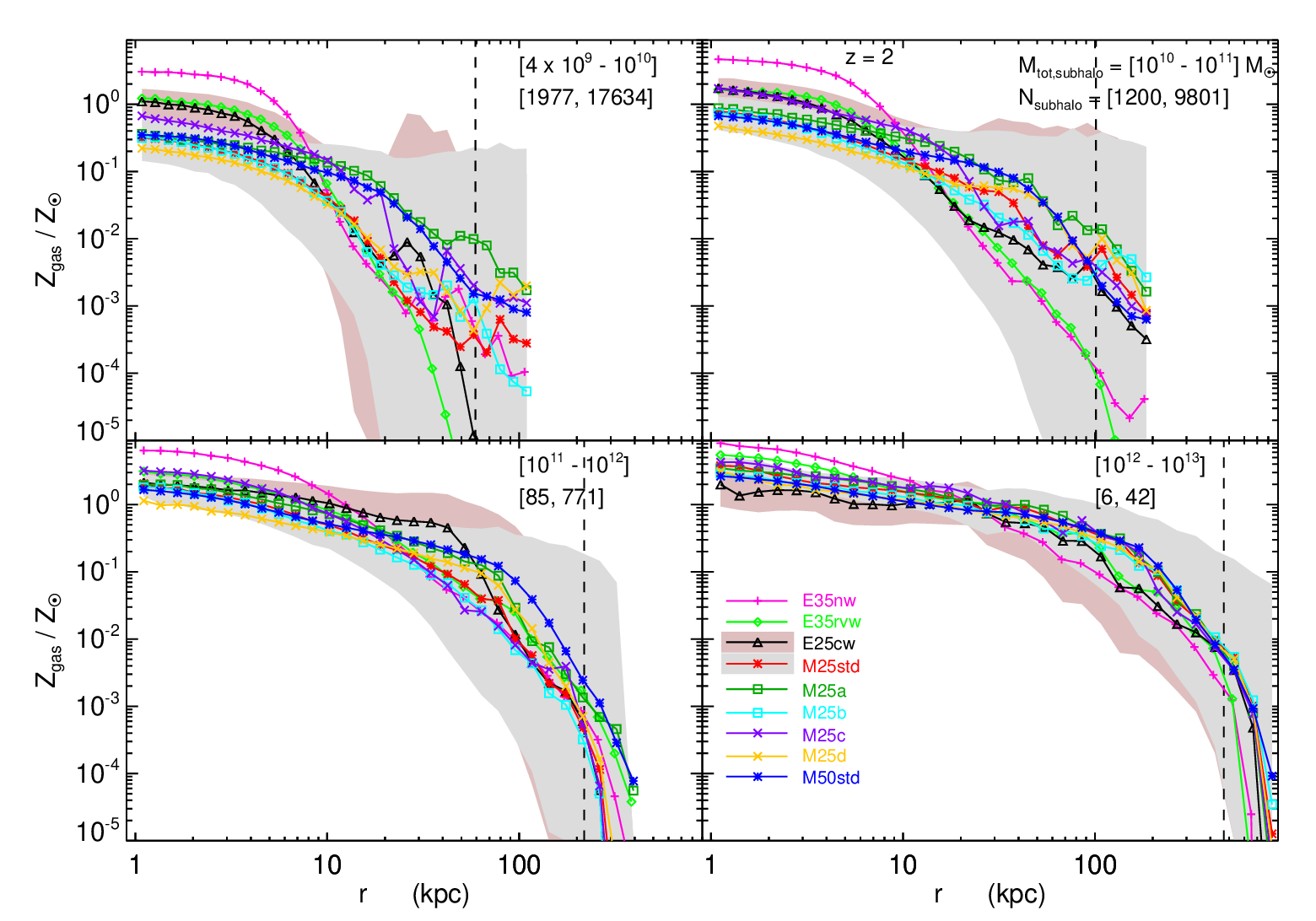} 
\caption{ 
Radial profiles of total gas metallicity around galaxy centers at $z = 2$, 
in a format similar to Fig.~\ref{fig-rho-vs-R-MassBins}. 
The ratio of all metal mass fraction in gas to that of the Sun ($Z_{\odot} = 0.0122$) is plotted. 
% The dashed curves represent median-$Z_{\rm gas}$ for the enriched particles only 
% in each radial bin, i.e. not counting particles with zero-$Z_{\rm gas}$ (\S\ref{sec-res-Radial-Ztot}). 
} 
\label{fig-Ztot-vs-R-MassBins} 
\end{figure*} 
%%%%%%%%%%%%%%%%%%%%%%%%%%%%%%%%%%%%%%%%%%%%%%%%%%%%%%%%%%%%%%%%%%%%%%%%  

Radial profiles of gas total metallicity ($Z_{\rm gas}$) are plotted 
in Fig.~\ref{fig-Ztot-vs-R-MassBins}, showing the ratio of mass fraction of all metals in the gas 
to that of the Sun, $Z_{\odot} = 0.0122$. 
The solid curve medians and shaded area percentile values are computed considering all 
(both enriched and non-enriched) gas particles in radial bins. 
% The dashed curves represent median-$Z_{\rm gas}$ for the enriched particles only, 
% i.e. those having $Z_{\rm gas} > 0$. 
% , without counting particles with $Z_{\rm gas} = 0$. 
% The dashed median $Z_{\rm gas}$ in Fig.~\ref{fig-Ztot-vs-R-MassBins} 
% is indistinguishable from the solid median $Z_{\rm gas}$ in the inner parts, 
% since most of the gas is enriched. 
% While at large-$r$ the dashed median $Z_{\rm gas}$ is higher, because the contribution 
% of the non-enriched ($Z_{\rm gas} = 0$) particles reduces the solid median $Z_{\rm gas}$, 
% as can be seen in all the four panels. 
% Between $R_{200} - R_{\rm lim}$, the enriched median $Z_{\rm gas}$'s are 
% $100 - 5000$ times higher than the total median $Z_{\rm gas}$'s. 
Some features of $Z_{\rm gas}$ are similar to the 
gas density profiles (\S\ref{sec-res-Radial-Density}), 
because metals are produced during SF which occurs in dense regions. 
At $r < (0.4 - 0.6) R_{200}$ all the profiles show decreasing $Z_{\rm gas}$ going outward 
from center, with varying $r$-dependent negative slopes. 
% The enriched-only $Z_{\rm gas}$ (dashed curves) rise again 
% from $r \geq (0.6 - 1) R_{200}$ and further outer regions. 
% It occurs because of a combination of reasons: 
% the presence of surrounding substructures where more metals are produced in-situ by ongoing SF, 
% and the spreading of metals by wind from the central SF regions into the CGM. 

All the MUPPI models produce relatively flatter metallicity profiles than the Effective models.  
Note that we do not use any metal loading factor in the sub-resolution recipe of any model. 
Gas particles carry away all of the metals they contain, and pollute the CGM. 
The no-wind run {\it E35nw}, which has no kinetic SN feedback, 
has the highest metallicity in the inner $r < (7 - 10)$ kpc. 
Wind feedback in the other runs suppresses central SF and transports some metal out, 
lowering the central $Z_{\rm gas}$. 
The trend reverses in the outer $r > (7 - 10)$ kpc: 
the kinetic feedback runs attain a higher $Z_{\rm gas}$ than {\it E35nw}, 
because of accumulation of metal-enriched gas expelled by wind. 
The differences are most prominent in the lower subhalo mass ranges 
($4 \times 10^{9} - 10^{10}$ and $10^{10} - 10^{11}$, top two panels), and decreases at higher masses. 

We infer that MUPPI distributes metals more adequately than the Effective models, 
which was already seen for a single galaxy in Fig.~\ref{fig-OneGal-Pos-Vel-rho-Z}, fourth row. 

% We infer from Fig.~\ref{fig-Ztot-vs-R-MassBins} that the CGM gas at galactocentric distances 
% close to and beyond $R_{200}$, within $r \sim (40 - 200)$ kpc comoving, 
% around galaxies of masses $M_{\rm subhalo} / M_{\odot} = 4 \times 10^{9} - 10^{11}$, 
% can give the best gas metallicity observational diagnostic 
% to distinguish between different galactic outflow models. 

%\subsubsection{Other Metals} 
%Take one run, show metallicity profile of different metals : C, Ca, O, N, Ne, Mg, S, Si, Fe. 

% \subsection{Fraction of Ejected (outflown) Gas from galaxies which Reaccreted at later epochs} 

\subsubsection{Comparison of Radial Profiles with Other Works} 
\label{sec-res-Radial-Compare} 

Performing zoom-in simulations of Milky-Way-mass disc galaxy formation, \citet{Hummels13} and 
\citet{Marinacci14} computed the radial profiles of diffuse CGM gas properties. 
Our density profiles (\S\ref{sec-res-Radial-Density}, Fig.~\ref{fig-rho-vs-R-MassBins}) and 
metallicity profiles (\S\ref{sec-res-Radial-Ztot}, Fig.~\ref{fig-Ztot-vs-R-MassBins}) 
are qualitatively similar to these studies. 
Our density profiles are also consistent to a work by \citet{Pallottini14}. 
Our $T$-profiles (\S\ref{sec-res-Radial-T}, bottom two panels of Fig.~\ref{fig-T-vs-R-MassBins}) 
present a local peak at $100 - 200$ kpc, which is at a larger-$r$ than the peak of \citet{Hummels13}. 
Our $Z_{\rm gas}$-profiles are qualitatively consistent with observations 
which show breaks (changes of slope) in the radial metallicity profiles, and/or 
rising metallicity gradients in the outer regions of galaxies \citep[e.g.,][]{Scarano12}. 

% the enriched-particle only median shown as dashed curves, which is analogous to mass-weighted metallicity  
% Hummels13 : Fig.~5 
% Marinacci14: Our density and metallicity profiles are qualitatively similar to their Figure 2 and 4. 
% Spherically-averaged density profiles of the diffuse gas as a function of radius 
% with the density profiles roughly composed of 3 sloped regions. 

\section{Summary and Conclusion} 
\label{sec-conclusion} 

% by numerically integrating a system of ODEs 
% (governing mass and energy exchange between hot, cold, and stellar phases) within the SPH timestep. 
% thus feedback is implemented using fully local properties % and analyse their impact on 

We quantify the properties of galactic outflows and diffuse gas over $z = 1 - 5$ 
by performing cosmological hydrodynamical simulations. 
We explore baryonic feedback models of star formation and SN feedback 
implemented within a modified version of the TreePM-SPH code {\sc GADGET-3}. 
Our novel sub-resolution model {\bf MUPPI} \citep{Murante10, Murante14} incorporates SF in multiphase ISM, 
and a direct distribution of thermal and kinetic energy from SN to the neighbouring gas, 
using the free parameters of feedback energy efficiency fraction and a probability. 
For comparison with MUPPI, we adopt the Effective SF model \citep{SH03} 
with variations of kinetic SN feedback, in the energy-driven formalism. 
Our simulations include additional sub-resolution physics: 
metal-dependent radiative cooling and heating in the presence of photoionizing background radiation; 
stellar evolution and chemical enrichment. 

We compare a total of nine simulations in this paper, done using the concordance $\Lambda$CDM model. 
We perform six new runs: five varying the SN feedback parameters of MUPPI, 
and one with the Effective model, 
of a $(25$ Mpc$)^3$ comoving cosmological volume with $2 \times 256^3$ DM and gas particles. 
An additional new MUPPI run is done of a larger $(50$ Mpc$)^3$ box 
with $2 \times 512^3$ particles, to increase the galaxy statistics. 
Two comparison simulations are taken from \citet{Barai13}: 
Effective SF model with no-wind and a radially varying wind case, 
which are $(35.56$ Mpc$)^3$ boxes simulated with $2 \times 320^3$ particles. 

Identifying galaxies using the SubFind halo finder, 
we measure the gas outflow of each galaxy by tracking the high-speed gas particles belonging to it. 
Only the central galaxies having a stellar mass higher than $10^{9} M_{\odot}$ are analysed. 
Our main findings are summarized below. 

\begin{itemize} 

\item[$\clubsuit$] The star formation rate density shows that at $z < 8$ 
the presence of kinetic SN feedback quenches SF more than any other feature of the SF model. 
The simulated SFRD has a plateau maximum between $z = 2 - 4$. 

% with a steeper reduction of SFRD on either side. % unlike the peak of observed SFRD at $z \sim 2.3$, 

% \item[-] {\it Gas kinematics.} 
% develop a substantial number of outflows: 
% The no-wind run presents outflowing gas only in galaxies with SFR $> 8 M_{\odot} /$ yr. 
% While the other runs have outflows in galaxies of all SFR ranges, as low as $0.1 M_{\odot} /$ yr. 

\item[$\clubsuit$] Kinetic SN feedback is able to drive gas and generate high-speed outflows 
in galaxies of all masses reached in our simulations, $M_{\rm halo} = (10^{10} - 10^{13}) M_{\odot}$. 
It also creates inflows via deceleration and later fall back of gas between $(1 - 2) R_{\rm gal}$. 
Both the outflows and inflows are heated in massive galaxy halos. 
We find the following trends at $z = 2$: 

\begin{itemize} 

\item[$\bullet$] When SN feedback is present, 
outflowing gas is detected in $(30 - 97) \%$ of the galaxies by number, 
depending on the model parameters and galaxy mass range. 
The number fraction of galaxies where outflow is detected increases with mass and SFR. 

\item[$\bullet$] Some outflow characteristics (listed next) exhibit positive correlations 
with galaxy halo, gas and stellar masses, as well as with the SFR, which is the tightest. 
However most of the cases present a large scatter. 

\begin{itemize} 

\item[$\spadesuit$] Measuring velocity of outflows by selecting gas above a fixed cutoff speed 
of $300$ km/s, the MUPPI model generates a positive correlation of $v_{\rm out}$ with galaxy SFR, 
while the Effective model shows a constant $v_{\rm out}$ with a large scatter. 

\item[$\spadesuit$] Mass outflow rate is measured by selecting gas above the escape velocity. 
The MUPPI model presents a stronger and relatively tighter positive correlation 
of $\dot{M}_{\rm out}$ with galaxy SFR. 
The Effective model runs with kinetic SN feedback exhibit a weak positive correlation of 
$\dot{M}_{\rm out}$ with SFR, with a slope flatter than the Muppi models; 
{\it E25cw} has a larger scatter, and {\it E35rvw} a tighter correlation. 
The no-wind run {\it E35nw} displays a scatter of $\dot{M}_{\rm out}$ having no relation with the SFR. 

\item[$\spadesuit$] The mass loading factor of the MUPPI outflows is constant with a scatter 
over all galaxy masses, 
$\eta = \dot{M}_{\rm out} / {\rm SFR} = 0.2 - 10$, and an average $\eta \sim 1$. 
The Effective models generate a negative correlation of $\eta$ with $M_{\rm halo}$. 
As an exception, run {\it E35rvw} presents a MUPPI-like trend at lower masses 
($M_{\rm halo} < 4 \times 10^{11} M_{\odot}$): 
$\eta$ is constant between $0.4 - 10$ versus $M_{\rm halo}$. 

% \item[$\spadesuit$] The overdensity of the outflows is constant with a scatter over the full range of SFR, 
% lying between $\delta = 200 - 3000$, and an average $\delta \sim 700$. 

% \item[$\spadesuit$] The MUPPI model and one Effective model run {\it E35rvw} 
% display a positive correlation of outflow temperature with galaxy SFR, rising upto 
% $T_{\rm gas} \sim 3 \times 10^{6}$ K at high-SFR. 
% % implying that the outflowing gas is heated to halo virial temperatures by shocks. 
% % The other two of the Effective models ({\it E35nw}, {\it E25cw}) present no correlation of temperature with SFR. 

% \item[$\spadesuit$] The outflow metallicity is scattered within 
% $Z_{\rm gas} = (0.01 - 4) Z_{\odot}$ in the Effective models, and 
% $Z_{\rm gas} = (0.2 - 4) Z_{\odot}$ in the MUPPI runs. 
% SN feedback in the MUPPI models is able to enrich the CGM of most galaxies to a relatively uniform value; 
% while the Effective models render some galaxies significantly less enriched. 

\end{itemize} 

Hence the MUPPI model, {\it using fully local properties} of gas as input to the sub-resolution recipe, 
is able to produce galactic outflows whose velocity and mass outflow rate 
{\it correlate with global properties of the galaxy (halo mass, SFR)}. 
This trend is found for the first time in cosmological simulations using sub-resolution models. 

The Effective model results are caused by the input fixed wind kick velocity 
in the energy-driven formalism. 
% and it reveals a weak pocorrelation of only $\dot{M}_{\rm out}$ with galaxy SFR. 

\item[$\bullet$] The shape of the outflows is inferred to be bi-polar in $\sim 95 \%$ MUPPI galaxies. 
If outflowing gas can escape the galaxy radius, 
in $\sim 90 \%$ cases it can escape the halo gravitational potential as well at the virial radius. 
The mass escape at the two radii are related as: 
$\dot{M}_{\rm out}(R_{\rm vir}) = 0.66 \dot{M}_{\rm out}(R_{\rm gal})^{0.98}$. 

% The fraction of galaxies where outflow is detected at $R_{\rm gal}$ 
% rises from $0.93$ to $0.97$ when using the bi-cylinder versus the sphere techniques, 
% which implies that 

% \item[$\bullet$] A warm ($T \sim 10^4 - 10^5$ K) and moderate-density ($\delta \sim 10^{2} - 10^{4}$) 
% gas phase occurs as outflow in the MUPPI model. 
% Such an outflow phase is absent in the Effective models. 

% gas overdensity radial profiles 
% The inner parts $r < 7 h^{-1}$ kpc of the two lower halo mass ranges 
% ($4 \times 10^{9} - 10^{10}$ and $10^{10} - 10^{11}$, top two panels) present notable differences: 
% {\it E35rvw} and {\it E25cw} runs produce a lower density, by $10 - 30$ times, than the others. 
% The relevant wind speed is independent of halo mass: constant $350$ km/s in {\it E25cw}, and 
% that dependent on galactocentric radius, $v_w(r)$, in {\it E35rvw}. 
% Such a velocity is high enough to eject the gas away from the halo potential in low-mass galaxies, 
% making the inner density smaller in these two runs. 
% 
% The central gas is not able to escape in cases {\it E35nw} (no-wind) which has no kinetic SN feedback, 
% and {\it M25c} which has less efficient feedback, 
% therefore producing the highest gas density at galaxy cores than the other runs. 

\item[$\bullet$] The gas temperature radial profiles reveal that 
the MUPPI model produces a colder ($T \sim 5 \times 10^{3} - 10^{5}$ K) galaxy core 
at $r \leq (3 - 8) h^{-1}$ kpc than the Effective model. 

% There is a change in $T$ slope in the outer parts, 
% at $r \geq 10 h^{-1}$ kpc in the top two panels, and 
% at $r \geq (60 - 100) h^{-1}$ kpc in the bottom two panels; 
% where the gas $T$ increases with radius, because of shock heating at galaxy outskirts. 

\item[$\bullet$] The MUPPI model generates relatively flatter gas metallicity radial profiles 
than the Effective model. 

% The no-wind run, has the highest metallicity in the inner $r < (5 - 7) h^{-1}$ kpc. 
% Wind feedback in the other runs suppresses central SF and transports some metal out, 
% lowering the central $Z_{\rm gas}$. 
% The trend reverses in the outer $r > (5 - 7) h^{-1}$ kpc: 
% the kinetic feedback runs attain a higher $Z_{\rm gas}$ than {\it E35nw}, 
% because of accumulation of metal-enriched gas expelled by wind. 
% over the full range of galaxy mass, earlier on 

\end{itemize} 

\item[$\clubsuit$] The fraction of MUPPI galaxies exhibiting an outflow are high at all times 
between $z = 1 - 5$, when outflow is measured by selecting gas above the escape velocity. 
The outflow detection fraction decreases gradually at lower redshifts 
from $f_{\rm outflow} = 0.99$ at $z = 5$, to $0.8$ at $z \sim 1$. 

The correlation between outflow velocity and SFR is positive at all the explored epochs: 
steeper at earlier times $z = 4 - 5$, and becomes flatter at later epochs. 
The mass loading factor is scattered within $\eta = 0.1 - 20$ between $z = 2 - 5$, 
and the average $\eta$ decreases with the passage of time. 
Later at $z = 0.8$, $\eta$ displays a negative correlation with $M_{\rm halo}$ at the high-mass end. 
The reason is the high SFR at $z = 2 - 4$ driving strong outflows in galaxies, 
while reduced SFR at later epochs quenches the outflow driving mechanism. 

\item[$\clubsuit$] Our results are overall consistent with observations of galactic winds. 
Galaxy population observations indicate that $v_{\rm out}$ increases with the SFR over $z = 0.7 - 2$, 
as we find in our simulations. 
Observations reveal that $z \sim 1 - 3$ galaxies almost ubiquitously possess powerful winds, 
than those at lower-$z$. 
This agrees with the decrease of outflow detection fraction in our simulations 
from $z = 3$ to later epochs. 

% The mass outflow rate (a few to 10's $M_{\odot} /$yr) and $\eta \sim 1$, 
% of single-galaxy detections at $z = 1 - 2.5$, lie within our simulation result range. 
% Our simulations predict a positive correlation of $v_{\rm out}$ with galaxy SFR 
% hence consistent with all these observations. 
% faster wind speeds at high-$z$ and slower winds at low-$z$. 

Our analysis demonstrates the ability of the MUPPI sub-resolution model 
to generate bi-polar outflows that present realistic properties. 
Additionally, we quantify the ability of both MUPPI and 
two variants of the standard energy-driven kinetic feedback model 
to produce significant outflows at $\sim 1 / 10$ of the virial radius and at the virial radius. 
Our study shows that, in the MUPPI model the fraction of energy really used to drive these outflows 
is only $\sim 1 / 3$ or less of that used by the code. 

\vspace{2 mm} 

As future work we would like to extract more observable statistics 
from the simulations. 
In particular we want to explore IGM metal-enrichment, by computing the 
Lyman-$\alpha$ flux and simulated quasar spectra, 
and compare them with observations of CGM and IGM at different impact parameters from galaxies. 
We also plan to peform better simulations in the future: 
run cosmological volumes with larger boxsize, and include AGN feedback in our models. 

\end{itemize}

\section*{Acknowledgements} 

We are most grateful to Volker Springel for allowing us to use the GADGET-3 code. 
We thank Stefano Borgani, Gabriella De Lucia, David Goz, Michaela Hirschmann, 
Edoardo Tescari, and Luca Tornatore, for useful discussions. 
The simulations were partly carried out at the CASPUR computing center with CPU time 
assigned under two standard grants. 
Post-processing was done on the local machine lapoderosa, 
and we acknowledge partial support from ``Consorzio per la Fisica - Trieste''. 
PB and MV acknowledge support from the ERC Starting Grant ``cosmoIGM'' and the INFN grant ``INDARK''. 
GM and PM acknowledge support from the PRIN-INAF 2012 grant 
``The Universe in a Box: Multi-scale Simulations of Cosmic Structures''. 
PM acknowledges a FRA2012 grant from the University of Trieste. 

% We thank Olga Cucciati for sending us observational data for SFRD, 
% and Edoardo Tescari for help with post-processing. 
% In addition, this work was supported by the Flagship Allocation Scheme of the NCI National Facility 
% at the Australian National University. 

%%%%%%%%%%%%%%%%%%%%%%%%%%%%%%%%%%%%%%%%%%%%%%%%%%%%%%%%%%%%%%%%%%%%%%%%
%
%                   REFERENCES
% \clearpage

%%%%%%%%%%%%%%%%%%%%%%%%%%%%%%%%%%%%%%%%%%%%%%%%%%%%%%%%%%%%%%%%%%%%%%%%%%%%%%%%%%%%%%%%%%
%%%%%%%%%%%%%%%%%%%%%%%%%%%%%%%%%%%%%%%%%%%%%%%%%%%%%%%%%%%%%%%%%%%%%%%%%%%%%%%%%%%%%%%%%%
%%%%%%%%%%%%%%%%%%%%%%%%%%%%%%%%%%%%%%%%%%%%%%%%%%%%%%%%%%%%%%%%%%%%%%%%%%%%%%%%%%%%%%%%%%

\end{document}